\documentclass[prd,aps,amsfonts,eqsecnum,superscriptaddress,nofootinbib,longbibliography,notitlepage]{revtex4-1}

\usepackage{graphicx}
\usepackage{xcolor}
\usepackage{rotating}
\usepackage{amsmath,amssymb,graphics,amsthm,isomath}
\usepackage{wasysym}

\usepackage[colorlinks=true, urlcolor=violet, linkcolor=blue, citecolor=red, hyperindex=true, linktocpage=true]{hyperref}
\usepackage[capitalise,compress]{cleveref}

\usepackage{caption}
\usepackage{subcaption}

\numberwithin{thm}{section}

\makeatletter
\renewcommand{\p@subsection}{}
\renewcommand{\p@subsubsection}{}
\makeatother

\usepackage{xcolor}
\usepackage{mathtools}

\usepackage{comment}

\def\bea{\begin{eqnarray}}
\def\eea{\end{eqnarray}}
\def\be{\begin{equation}}
\def\ee{\end{equation}}
\def\bes{\begin{subequations}}
\def\ees{\end{subequations}}
\def\bed{\begin{displaymath}}
\def\eed{\end{displaymath}}
\def\beal{\begin{aligned}}
\def\eeal{\end{aligned}}
\def\bew{\begin{widetext}}
\def\eew{\end{widetext}}
\def\beit{\begin{itemize}}
\def\eeit{\end{itemize}}
\def\bea{\begin{array}}
\def\eea{\end{array}}
\def\been{\begin{enumerate}}
\def\eeen{\end{enumerate}}

\makeatletter
\DeclareRobustCommand{\cev}[1]{%
  \mathpalette\do@cev{#1}%
}
\newcommand{\do@cev}[2]{%
  \fix@cev{#1}{+}%
  \reflectbox{$\m@th#1\vec{\reflectbox{$\fix@cev{#1}{-}\m@th#1#2\fix@cev{#1}{+}$}}$}%
  \fix@cev{#1}{-}%
}
\newcommand{\fix@cev}[2]{%
  \ifx#1\displaystyle
    \mkern#23mu
  \else
    \ifx#1\textstyle
      \mkern#23mu
    \else
      \ifx#1\scriptstyle
        \mkern#22mu
      \else
        \mkern#22mu
      \fi
    \fi
  \fi
}

\makeatother

\newcommand{\lee}{\ell_{\mathrm{ee}}}
\newcommand{\lei}{\ell_{\mathrm{ei}}}

\usepackage{dsfont}

\begin{document}

\title{Distinguishing viscous, ballistic, and diffusive current flows in anisotropic metals}

\author{Marvin Qi}
\email{marvin.qi@colorado.edu}
\affiliation{Department of Physics and Center for Theory of Quantum Matter, University of Colorado, Boulder CO 80309, USA}

\author{Andrew Lucas}
\email{andrew.j.lucas@colorado.edu}
\affiliation{Department of Physics and Center for Theory of Quantum Matter, University of Colorado, Boulder CO 80309, USA}

\begin{abstract}
We show that in anisotropic Fermi liquids where momentum-conserving  scattering is much faster than momentum-relaxing scattering processes, local imaging of the electric current flow patterns can cleanly distinguish between ballistic, viscous and ohmic flow patterns simultaneously (using a single image).    We propose using multi-layer graphene-based heterostructures, including ABA trilayer graphene, as a natural experimental platform where an anisotropic ballistic-to-viscous crossover may be visible in near-term experiments.  Such experiments could lead to more direct measurements of momentum-conserving scattering lengths in electronic Fermi liquids.
\end{abstract}

\date{\today}

\maketitle
\tableofcontents

\section{Introduction}
After preliminary work decades ago \cite{gurzhi,de_Jong_1995}, the past few years have seen intense experimental efforts \cite{Crossno_2016,Bandurin_2016,Moll_2016,Krishna_Kumar_2017,ghahari,Gallagher2019,Berdyugin_2019,Gooth_2018,Gusev_2018} to study the emergence of viscous hydrodynamics in electron liquids in high quality metals;  see \cite{Andreev2011,Forcella_2014,pellegrino,Tomadin_2014,Torre_2015,Levitov_2016,agarwal,Guo_2017,alekseev} for previous theoretical work, and \cite{Lucas_2018, Narozhny_2017} for reviews.   One fundamental motivation for studying hydrodynamic phenomena is that they are relatively universal from one material to the next, dependent only on  crystalline symmetries \cite{caleb,varnavides2020generalized,cook2021viscometry} but not on peculiarities of the scattering mechanisms.   Moreover, measuring hydrodynamic coefficients gives access to momentum-conserving scattering mechanisms that are invisible in ordinary bulk resistivity measurements (which, in the absence of umklapp processes, are sensitive to impurities and momentum-relaxing processes).   As a consequence, a better understanding of electronic hydrodynamics will help us to characterize the dynamical processes that take place in metals, especially those with strongly correlated electrons.

In a Fermi liquid of electrons in a metal, we may talk about a mean free path for both impurity (or more generic momentum-relaxing) scattering processes ($\lei$), together with a mean free path for momentum-conserving collisions ($\lee$), which are typically electron-electron collisions.\footnote{We note in passing, however, that it is also possible in principle for phonon scattering to be momentum-conserving \cite{gurzhi1972,levchenko2020transport,xiaoyang1}.}  On length scales $L \gg \lei$, transport will be dominated by impurity scattering, and the physics is consistent with a local form of Ohm's Law ($J_i = \sigma E_i$), with $\sigma$ a constant.   However, on length scales $L \ll \lei$, it may be the case that $\sigma$ is \emph{not} simply a constant, but rather a non-trivial function, obeying a ``generalized" Ohm's Law (linear response relation) \begin{equation}
    J_i(x) = \int \mathrm{d}^dx^\prime \; \sigma_{ij}(x-x^\prime)E_j(x^\prime), \label{eq:nonlocalohm}
\end{equation}
where we are here using the Einstein summation convention on repeated indices.  By studying transport on various length scales $L$, one might hope to resolve not simply a single constant $\sigma$ (the bulk conductivity), but rather information about a function $\sigma_{ij}(x-x^\prime)$.

An exciting new experimental development over the past few years has been the application of imaging methods to directly and non-invasively image hydrodynamic flows in materials.  In some experiments \cite{sulpizio,krebs2021imaging}, this is achieved by using local potentiometry to study the local voltage in the device, while in others a nitrogen-vacancy-center magnetometer \cite{Ku_2020,jenkins2020imaging,vool2020imaging} is used to image current flow patterns.  In both cases, evidence can be obtained for hydrodynamic flow by studying the spatial patterns imaged, along with their temperature dependence.  These images can be combined with theoretical models to extract $\sigma_{ij}(x-x^\prime)$ and thus reveal more transport physics than had previously been accessible.

However, in most of the experiments done thus far to image electron hydrodynamics, the distinctions between different transport regimes are more quantitative than qualitative:   e.g. in Poiseuille flow through a channel, the current can flow relatively uniformly through the device \emph{either} when $L \ll \lee$ \emph{or} $L\gg \lei$ \cite{Ku_2020}.  Moreover, it is highly desirable to unambiguously distinguish between three transport regimes \cite{Lucas_2018}: ballistic ($L\lesssim \lee$), viscous ($\lee \lesssim L\lesssim \sqrt{\lee\lei}$),  and \emph{ohmic} ($L \gtrsim \sqrt{\lee \lei}$).  The Gurzhi length $\sqrt{\lee \lei}$ can be understood as coming from a ``random walk", whereby a quasiparticle carrying some transverse momentum undergoes a random walk with diffusion constant $D\sim v_{\mathrm{F}}\lee$ for a time $t\sim \lei/v_{\mathrm{F}}$, when it hits an impurity and loses momentum: thus propagating a typical distance $\sqrt{Dt}\sim \sqrt{\lee \lei}$.

The purpose of this paper is to explain how to cleanly achieve these capabilities using magnetometry in an anisotropic Fermi liquid, where we will show that even a single image of current flow patterns can be sufficient to easily distinguish between the three transport regimes above.  We will primarily do so by studying a microscopically ``realistic" toy model of a two-dimensional electron fluid that arises in ABA-trilayer graphene \cite{newdiracpoints,abatrilayer}, a system with highly tunable anisotropic Fermi surfaces \emph{and} a low carrier density:  properties which make this platform ideal for anisotropic electron hydrodynamics.   However, we will also show that all of the qualitative features of that system are also found in other toy models of anisotropic Fermi liquids.  Hence we expect that the methods we propose here are a natural way to systematically and more effectively use imaging techniques to measure the fundamental length scales governing electronic dynamics.  

\section{General framework}\label{sec:method}
We begin this paper by describing an ``all purpose" algorithm for predicting the flow of current through a complex device.  This method was first used in \cite{Guo_2017}, and later generalized in \cite{jenkins2020imaging,huang2021fingerprints}.  

\subsection{Algorithm for generating flow patterns}

\begin{figure}[t]
\centering
\includegraphics[width=3in]{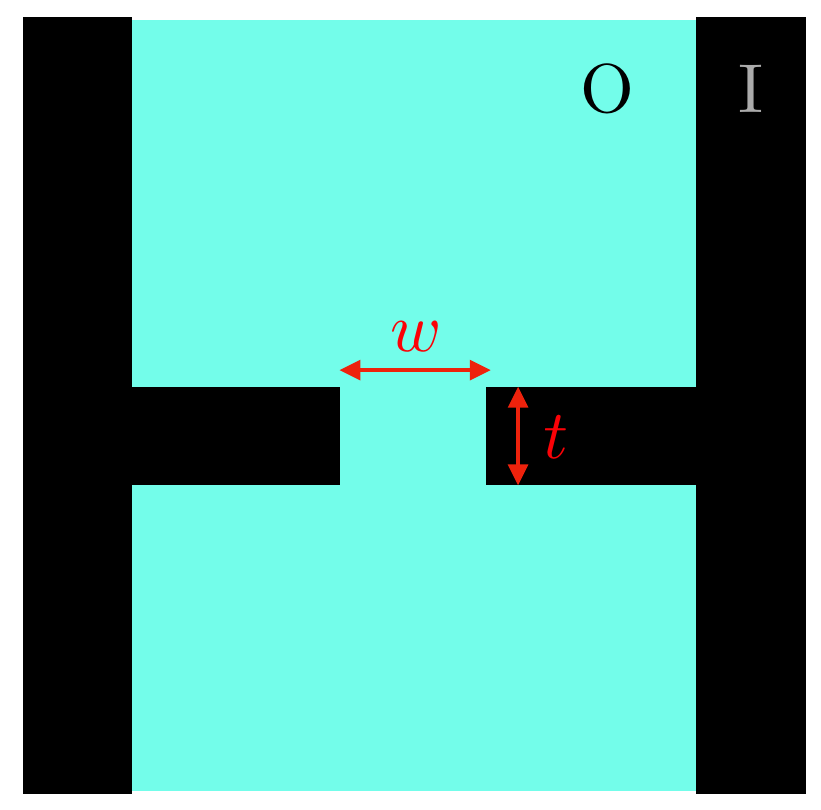}
\caption{A sketch of the constriction geometry.  Current can flow in the O (outside) light blue region, and not in the I (inside) black region.  The specific constriction of interest has a thickness $t$ and a width $w$ across.}
\label{fig:constriction}
\end{figure}

We will consider a geometry of the form sketched in Figure \ref{fig:constriction}.   While this figure shows a particularly interesting constriction geometry for experiments, the formalism we describe below is more general as well.  For experimental purposes however, we will restrict our focus to a two-dimensional theory, as in two-dimensions it is natural to use an out-of-plane scanning probe to study the local physics non-invasively.\footnote{We also hope, however, that these methods will apply effectively to thin films of otherwise three-dimensional crystals as well).}  As depicted in Figure \ref{fig:constriction}, we divide space into two separate regions:  I (inside) and O (outside) the insulating constriction.  Within the theory of linear response, we can ``formally" write (\ref{eq:nonlocalohm}).  We make the ansatz that \begin{equation}
E_i(x) = E_i^{\mathrm{ext}} + E_j^{\mathrm{ind}}(x), 
\end{equation}
where $E_j^{\mathrm{ext}}$ is a uniform, externally applied,  electric field which drives current through the device, and $E_j^{\mathrm{ind}}(x)$ is the spatially induced electric field caused by the presence of the constriction.   In a general many-body system, \cite{hartnoll2016holographic} \begin{equation}
\sigma_{ij}(x,x^\prime) = -\mathrm{i} \left. \frac{\partial G^{\mathrm{R}}_{J_i J_j}(x,x^\prime,\omega)}{\partial \omega} \right|_{\omega\rightarrow 0}, \label{eq:sigmaGR}
\end{equation}   
with $x$ and $x^\prime$ denoting position vectors in $d$ spatial dimensions (in this paper, $d=2$ always), and the retarded Green's function is defined (up to a Fourier transform) via \begin{equation}
    G^{\mathrm{R}}_{J_iJ_j}(x,x^\prime,t) := \mathrm{i}\mathrm{\Theta}(t) \langle [J_i(x,t),J_j(x^\prime,0)]\rangle.
\end{equation}

 Thus far, we have arguably just replaced one hard problem (finding $J_i$) with another one (finding $E_i^{\mathrm{ind}}$). The key simplification in this re-packaging, however, is that it is easier to make an ``educated guess" about the form of $E_i^{\mathrm{ind}}$ -- effectively amounting to a choice of boundary conditions.  While in ordinary physical problems (such as hydrodynamic or diffusive partial differential equations) one typically demands sufficiently many boundary conditions be obeyed (which are posited at the start of the solution), and then looks for the exact solution that obeys those boundary conditions,  here we take a slightly distinct perspective.  Since the equation (\ref{eq:nonlocalohm}) is a non-local integral equation, it seems easier to approach this problem not by demanding integral boundary conditions (which we do not have a good \emph{ab initio} theory for reagrdless), but instead by simply positing a particularly simple form of $E_i^{\mathrm{ind}}$ that is not manifestly unphysical. 
 
 To motivate a clever choice of $E_i^{\mathrm{ind}}$, we proceed as follows. It seems natural that $E^{\mathrm{ind}}_i$ is only non-vanishing in region I.  Moreover, since current cannot flow through region I, we must have \begin{equation}
0  = \sigma_{ij}(\mathbf{k}=\mathbf{0}) E_j^{\mathrm{ext}} + \int_{\mathrm{I}} \mathrm{d}^2y \sigma_{ij}(x-y) E_j^{\mathrm{ind}}(y). \label{eq:EIeq}
\end{equation}
Here $\sigma_{ij}(\mathbf{k}=\mathbf{0})$ is the homogeneous (zero wave number) part of the conductivity.  While solutions to this equation may not be unique, one solution to this equation is \begin{equation}
E_i^{\mathrm{ind}}(x) = - \int\limits_{\mathrm{I}} \mathrm{d}^2y \sigma_{\mathrm{II},ij}^{-1}(x,y)) \sigma_{jk}(\mathbf{k}=\mathbf{0}) E_k^{\mathrm{ext}} \label{eq:EiI}
\end{equation}
where $\sigma_{\mathrm{II},ij}(x,y)$ is the sub-block of the matrix $\sigma_{ij}(x-y)$ when both $x$ and $y$ lie in the domain I.   Since $\sigma_{jk}(x-y)$ has an infinite number of zero modes due to the fact that $\sigma_{ij}$ projects off all longitudinal spatially variable currents -- see the later formula (\ref{eq:Sigma}) -- in practice we numerically add a tiny regulator to $\sigma_{\mathrm{II}}$ prior to inverting it: \begin{equation}
    \sigma_{\mathrm{II},ij}^{-1}(x,y) \approx (\sigma_{\mathrm{II},ij}(x-y) + b\delta(x-y))^{-1}, \;\; (b\rightarrow 0).
\end{equation}   
Then, we simply plug in $E_i^{\mathrm{ind}}$ from (\ref{eq:EiI}) into (\ref{eq:nonlocalohm}) to evaluate the electric current in the domain O outside the constriction. It is numerically efficient to perform these computations, as the only matrix inverse involves the degrees of freedom inside the constriction, which can be a small subset of the grid points in the total domain.  

In typical numerical simulations, we take around 50-250 grid points in the $x$ and $y$ directions, while leaving the constriction to correspond to around 5\% of the overall grid points in the system.  This means that the matrix inverse $\sigma_{\mathrm{II}}^{-1}$ can be achieved by inverting a matrix with $\sim 2000$ rows and columns, which is easily done on a laptop computer.

Having described our method, let us now return to its justification. In some sense, this method corresponds to choosing boundary conditions that make the integral equation have a particularly ``simple" solution.  We believe that this is the best justification for the approach.  As an analogy, consider a textbook Navier-Stokes system, where two historically popular choices of boundary conditions are either ``no slip" (fluid velocity $v=0$ at the boundary) or ``no stress" (fluid velocity normal to boundary $v_n=0$, while perpendicular components have vanishing normal derivative $\partial_n v_\parallel = 0$).  If one interpolates between these two boundary conditions by choosing $(1+\lambda \partial_n)v_\parallel = 0 $, then the large-scale qualitative properties of the flow patterns are the same for finite $\lambda$, while being special at $\lambda=\infty$.   A randomly chosen value of $\lambda$ will thus lead to a qualitatively similar flow pattern, with deviations from the $\lambda=0$ flow at distances $\lesssim \lambda$ away from boundaries.

We emphasize that highly symmetric geometries, such as flows down an (infinitely) long channel, are highly sensitive to boundary conditions \cite{Lucas_2018}; as such, our algorithm is likely better when applied to more complex scenarios where ``slip" boundary conditions are not the only important ones in the problem.  Of course, even in a symmetric set-up, we can generate \emph{one} solution to microscopic linear response theory using this algorithm.  We do note, however, that even our constriction geometry has recently been shown to have some sensitive to boundary conditions \cite{Pershoguba2020}: flow patterns appear different when $\lambda=\infty$ vs. $\lambda=0$ (in the language of the slip length $\lambda$, introduced above).  As discussed there, however, we anticipate that our algorithm (which gives flow patterns closer to $\lambda=0$) represents the more ``generic" result, and indeed our approach is quantitatively in agreement with experiments \cite{jenkins2020imaging}.

\subsection{Some useful identities}

In the remainder of this section, we state a few useful properties about $\sigma_{ij}(k)$ which are quite general, and reduce the difficulty of theoretical computations of the conductivity.  We begin by noting that the current must be conserved: \begin{equation}
\partial_i J_i = 0. \label{eq:continuity}
\end{equation} Taking the Fourier transform of (\ref{eq:nonlocalohm}), we find that (\ref{eq:continuity}) and the symmetries of transport coefficients imply that \begin{equation}
\sigma_{ij}(k) = \left(\delta_{ij}  - \frac{k_ik_j}{k^2}\right)\Sigma (k),   \;\;\;\;  (\mathbf{k} \ne \mathbf{0}).  \label{eq:Sigma}
\end{equation}
in all models that are either time-reversal symmetric, or in which time-reversal symmetry is broken but preserved in conjuction with another operation (e.g., a magnetic field preserves the combination of parity and time-reversal symmetries).  To understand why, first assume time reversal symmetry; by Onsager reciprocity $\sigma_{ij}(k)$ is symmetric.   The projector removes 3 out of 4 possible matrix elements of $\sigma_{ij}$; all that is left is a single component, whose magnitude we define as $\Sigma(k)$.   In a background magnetic field, time reversal symmetry is broken, but Onsager reciprocity still allows us to show that if one row of $\sigma_{ij}$ vanishes for any $B$, so does one column; hence, just as before, we obtain the simplified form (\ref{eq:Sigma}).  Note that when $\mathbf{k}=\mathbf{0}$, this argument fails and in general all 4 components of $\sigma_{ij}$ could be distinct (without time-reversal symmetry).


From the form of (\ref{eq:sigmaGR}), it is also clear how to generalize the formulae above to cases where one wants to image the spatial patterns of operators besides current flow.  For example, using scanning SETs, charge density has been imaged in a recent experiment to look for hydrodynamics \cite{sulpizio}.  If we want to image the operator $\mathcal{O}(\mathbf{x})$, then in terms of the generalized conductivity \begin{equation}
    \sigma_{\mathcal{O},i}(x-y) = -\mathrm{i}  \left. \frac{\partial G^{\mathrm{R}}_{\mathcal{O} J_i}(x,y,\omega)}{\partial \omega} \right|_{\omega\rightarrow 0},
\end{equation}
we find that our arguments generalize to\begin{equation}
    \langle\mathcal{O}(x)\rangle = \sigma_{\mathcal{O},i}(\mathbf{k}=\mathbf{0})E_i -  \int \mathrm{d}^2y \; \sigma_{\mathcal{O}i}(x-y) \sigma^{-1}_{\mathrm{II},ij}(x,y) \sigma_{jk}(\mathbf{k}=\mathbf{0})E_k.
\end{equation}
In this paper, we will focus on the case where $\mathcal{O}=J_i$ is a current operator itself.

\section{Kinetic theory} \label{sec:kinetictheory}
Our proposed algorithm has a number of substantial advantages for modeling transport in complex geometries and in complex materials.  For example, consider obtaining $\Sigma(k)$ by solving a linearized quantum Boltzmann equation.  As we will argue below, this is actually quite tractable, even when employing microscopic band structure theory to model the Fermi surfaces found in particular materials.  (For recent work that also aims to compute the collision integrals from first principles -- an important but much more difficult and computationally expensive task -- see \cite{coulter,varnavides2021finitesize}.)  The computation of $\Sigma(k)$ will generically involve a numerical computation within the Brillouin zone; however, the computation only takes place in the two dimensions of $p$-space.  In contrast, a direct solution of the Boltzmann equation in the complex geometry is a four dimensional integrodifferential equation in both position $x$ and momentum $p$, moreover being non-local in the latter two dimensions!  

We now briefly describe how to combine the algorithm of the previous section with a toy model for the quantum Boltzmann equation in the relaxation time approximation, to generate a somewhat ``ab initio" model for non-local transport that can be efficiently solved on a laptop computer.   We consider a two-dimensional Fermi liquid at very low temperature, in a regime where it is appropriate to approximate that charge and momentum are the only conserved quantities.\footnote{Outside of studying transport in inhomogeneous disordered metals \cite{Andreev2011}, this is a reasonable assumption \cite{Lucas_2018}.}   Following the notation and framework described in \cite{lucashartnoll}, we will study the linear response regime of the Boltzmann equation governing the distribution function $f(x,p)$ of a weakly interacting Fermi liquid with dispersion relation $\epsilon(p)$ (we neglect spin).  Weakly perturbing away from equilibrium, we may write \begin{equation}
f(x,p) = f_{\mathrm{eq}}(\epsilon(p)) - \frac{\partial f_{\mathrm{eq}}}{\partial \epsilon} \Phi(x,p)
\end{equation}
where $f_{\mathrm{eq}}$ is the equilibrium Fermi-Dirac distribution: \begin{equation}
f_{\mathrm{eq}}(\epsilon) = \frac{1}{1+\mathrm{e}^{\epsilon/T}}.
\end{equation}
The function $\Phi$ will be smooth and better behaved in $p$-space than $f-f_{\mathrm{eq}}$ which is quite sharply peaked near the Fermi surface.  

We then define a bra-ket notation which allows us to re-cast kinetic equations in terms of linear algebra problems.  We write \begin{equation}\label{eq:vectordef}
|\Phi(x)\rangle = \int\mathrm{d}^2p \; \Phi(x,p)|p\rangle
\end{equation}
along with the inner product \begin{equation}
\langle p|p^\prime\rangle = - \frac{\partial f_{\mathrm{eq}}}{\partial \epsilon}\frac{\delta(p-p^\prime)}{(2\pi\hbar)^2} .
\end{equation}
Upon taking a Fourier transform in the $x$ direction, the time-dependent Boltzmann equation is
\begin{equation}
\partial_t |\Phi\rangle + \mathrm{i} k \cdot v(p) |\Phi\rangle + \mathsf{W}|\Phi\rangle = \partial_t |\Phi\rangle +\mathsf{G}^{-1}|\Phi\rangle
\end{equation}
where $v(p) = \partial \epsilon/\partial p$,  $\mathsf{W}$ is the linearized collision integral, and $\mathsf{G}$ is defined above through $\mathsf{G}^{-1}=\mathsf{W}+\mathrm{i}k\cdot v$.
In this framework, one can show that \cite{lucashartnoll}
\begin{equation}
\sigma_{ij}(k) = \langle \mathsf{J}_i| \mathsf{G}(k) |\mathsf{J}_j\rangle, \label{eq:sigmaijkinetic}
\end{equation}
where we have defined $|\mathsf{J}_i \rangle = \int \mathrm{d}^2p\; v_i |p\rangle$ as per Eq. \ref{eq:vectordef}.

In general, finding the linearized collision integral $\mathsf{W}$ is not trivial.  However, using a relaxation time approximation, we can dramatically simplify the computation of $\mathsf{G}$, following \cite{Guo_2017}. We first define a matrix $\mathsf{G}_0$, which corresponds to a simple kinetic equation where $f(x,p)$ relaxes to 0 at the rate $\gamma_{\mathrm{ee}}$:
\begin{equation}
\mathsf{G}_0^{-1} |p\rangle = (\gamma_{\mathrm{ee}} + \mathrm{i}k\cdot v(p))|p\rangle.
\end{equation}
The notation $\gamma_{\mathrm{ee}}$ is meant to emphasize that this relaxation time is associated with an electron-electron scattering rate, which will be the fast decay time for most modes in our final theory.   Unfortunately, this $\mathsf{G}_0^{-1}$ does not have \emph{any} conserved quantities, yet charge must be exactly conserved, and momentum may be long-lived.  So, defining the vectors \begin{subequations}\begin{align}
|\mathsf{n}\rangle &= \int\mathrm{d}^2p \; |p\rangle , \\
|\mathsf{P}_i\rangle &= \int\mathrm{d}^2p \;  p_i |p\rangle , 
\end{align}\end{subequations}
we can obtain the true $\mathsf{G}$ of our relaxation time model by modifying $\mathsf{G}_0$ in order to account for the slower relaxation times of density and momentum:
 \begin{equation}
\mathsf{G}^{-1} = \mathsf{G}_0^{-1} - \mathsf{PRP}
\end{equation}
where \begin{equation}\label{eq:projector}
\mathsf{P} = \frac{|\mathsf{n}\rangle\langle\mathsf{n}|}{\langle\mathsf{n}|\mathsf{n}\rangle} + \frac{|\mathsf{P}_x\rangle\langle\mathsf{P}_x|}{\langle\mathsf{P_x}|\mathsf{P}_x\rangle} + \frac{|\mathsf{P}_y\rangle\langle\mathsf{P}_y|}{\langle\mathsf{P}_y|\mathsf{P}_y\rangle} 
\end{equation}
(in this relation, we have assumed that the three vectors above are orthogonal, which holds for all models considered in this work) and $\mathsf{R}$ encodes the \emph{difference in} relaxation rates for the various conservation laws;  we take \begin{equation}
\mathsf{R} =  \gamma_{\mathrm{ee}} |\mathsf{n}\rangle\langle\mathsf{n}|+ (\gamma_{\mathrm{ee}}-\gamma_{\mathrm{ei}}) |\mathsf{P}_x\rangle\langle\mathsf{P}_x| +(\gamma_{\mathrm{ee}}-\gamma_{\mathrm{ei}}) |\mathsf{P}_y\rangle\langle\mathsf{P}_y|
\end{equation}
Here $\gamma_{\mathrm{ei}}$ denotes the electron-impurity scattering rate, which must obey $\gamma_{\mathrm{ee}} \ge \gamma_{\mathrm{ei}}$.    

Observe that $\mathsf{G}$ and $\mathsf{G}_0$ are infinite dimensional matrices, but since $\mathsf{G}_0^{-1}$ is diagonal, it is trivial to evaluate $\mathsf{G}_0$.   Using the identity \begin{equation}
\mathsf{G} = \mathsf{G}_0 + \mathsf{G}_0\mathsf{PR}(1-\mathsf{PG}_0\mathsf{PR})^{-1}\mathsf{PG}_0, \label{eq:GG0}
\end{equation}
we find a simple formula for $\mathsf{G}_0$ that involves the inverse of only a $3\times 3$ matrix: by construction, $(1-\mathsf{PG}_0\mathsf{PR})$ is a trivial matrix outside of the 3-dimensional subspace of slow (hydrodynamic) modes that  $\mathsf{P}$ projects onto.   Therefore, we can easily numerically evaluate $\mathsf{G}$ for arbitrary $\epsilon(\mathbf{p})$ without inverting any large matrices -- the computational cost primarily comes from simply evaluating the inner products such as $\langle \mathsf{J}_i|\mathsf{P}_j\rangle$, which involve numerically evaluated integrals over the Brillouin zone.  Indeed, combining (\ref{eq:Sigma}), (\ref{eq:sigmaijkinetic}) and (\ref{eq:GG0}), we arrive at the desired result: \begin{equation}
\Sigma(\mathbf{k}) = \langle \mathsf{J}_x | \mathsf{G}(\mathbf{k})|\mathsf{J}_x\rangle  + \langle \mathsf{J}_y | \mathsf{G}(\mathbf{k})|\mathsf{J}_y\rangle . \label{eq:sigmakinetic}
\end{equation}

In order to numerically evaluate these inner products, especially given a dispersion relation that may be numerically calculated itself (as for ABA graphene below), we typically consider a finite $T=T_*$ in the algorithm above, such that the numerical discretization of the Brillouin zone has a spacing $\mathrm{\Delta}p$ obeying \begin{equation}
    v_{\mathrm{F}} \mathrm{\Delta}p \lesssim T_*.
\end{equation}
A standard Riemann sum approximation to the integrals defined by the inner product may then be employed.  In this formula, $v_{\mathrm{F}}$ denotes the magnitude of the quasiparticle velocity at that point on the Fermi surface, and it can vary as we move along the Fermi surface.

\section{ABA-trilayer graphene}
In this section, we will evaluate (\ref{eq:sigmakinetic}) explicitly within a microscopic band structure theory of ABA-trilayer graphene \cite{newdiracpoints,abatrilayer}. ABA graphene has multiple features which make it interesting to study in the context of looking for hydrodynamics in an anisotropic electron liquid. Firstly, it has been studied in previous experiments \cite{abatrilayer}.  Secondly, we may anticipate that it is a reasonably high-quality material with low impurity concentrations and thus small momentum-relaxing scattering rates: indeed, in monolayer or bilayer graphene, the momentum-relaxing mean free paths are multiple microns at low temperatures due to the efficient growth of van der Waals heterostructures which encapsulate the graphene layers with insulators such as hexagonal boron nitride (hBN) \cite{Dean_2010}.   At higher temperatures, the weak electron-phonon coupling between graphene and hBN leads to stronger momentum relaxation, and we expect it is unlikely for things to drastically differ for ABA graphene.  (Unfortunately, we were unable to find transport characterizations of ABA graphene in the experimental literature.)  Thirdly, monolayer and bilayer graphene are widely observed to have much faster momentum-conserving scattering rates, relative to momentum-relaxing rates, and thus give rise to some form of approximate viscous hydrodynamics, in experiment \cite{Krishna_Kumar_2017,sulpizio,jenkins2020imaging,krebs2021imaging}:  since we will show that ABA graphene also has the low density and small Fermi surfaces which enable rapid momentum-conserving scattering, we anticipate that this platform will also give rise to a viscous electron fluid.   Lastly and most importantly for us, ABA-trilayer graphene exhibits highly tunable anisotropic Fermi surfaces, which are essential to crisply distinguish the viscous from ballistic current flows.  These Fermi surfaces can be tuned by varying the charge density in the device alone, which is easy to accomplish in a van der Waals heterostructure.  

\subsection{Band structure}
\begin{figure}
    \centering
    \begin{subfigure}[b]{0.45\textwidth}
        \centering
        \includegraphics[width=\textwidth]{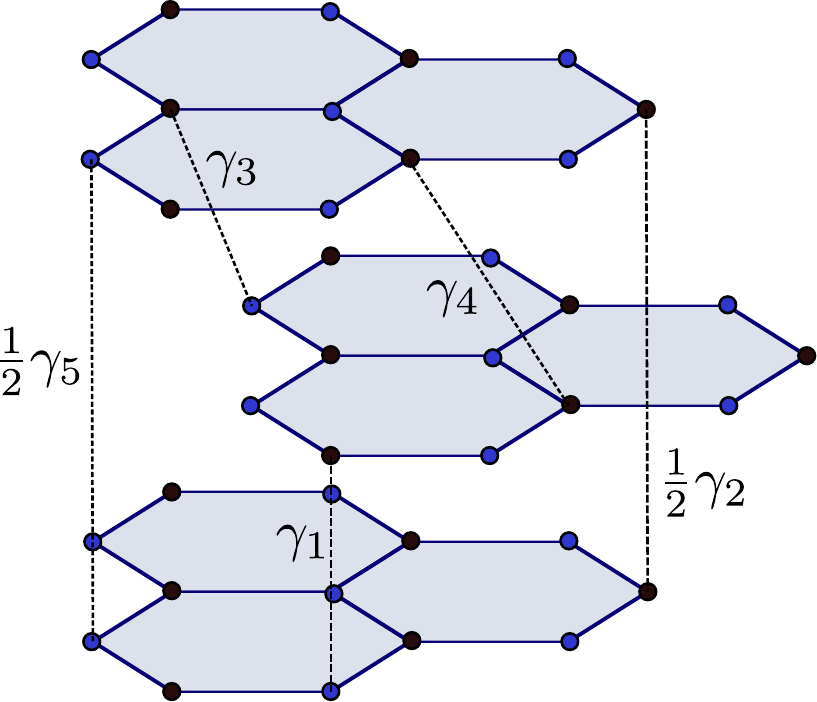}
        \label{fig:abastack}
    \end{subfigure}
    \begin{subfigure}[b]{0.45\textwidth}
        \centering
        \includegraphics[width=\textwidth]{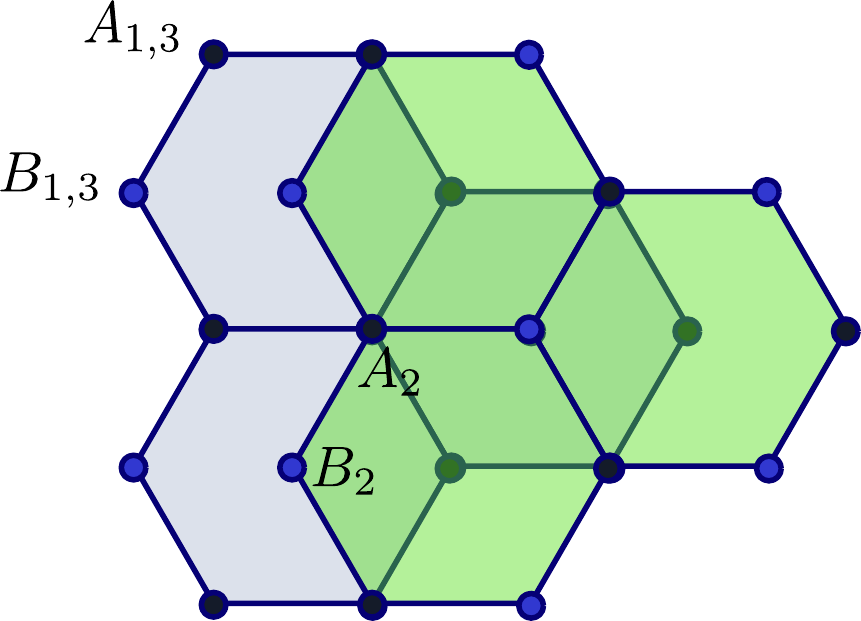}
        \label{fig:abatopview}
    \end{subfigure}
    \caption{The stacking geometry of ABA trilayer graphene, viewed from the side and from the top. The $A$ sublattice (black) of the middle layer is aligned with the $B$ sublattices (blue) of the outer two layers. The hopping parameters in (\ref{eq:hopping}) are shown. }
    \label{fig:aba}
\end{figure}

Having motivated our interest in studying transport in this material, we now describe a microscopic model for the band structure of ABA-trilayer graphene, adopted from \cite{newdiracpoints}.   Recall that single layer graphene consists of a honeycomb lattice of carbon atoms.  The honeycomb lattice is bipartite, and consists of $A$ and $B$ sublattices.  As shown in Figure \ref{fig:aba}, 
ABA-trilayer graphene consists of three graphene layers stacked such that the $A$ sublattice of the middle layer overlaps with the $B$ sublattice of the outer two layers. Each unit cell therefore consists of the $A_i, B_i$ sublattices of each layer $i \in \{1,2,3\}$.   Our tight-binding model can then be written as a $6 \times 6$ Hermitian matrix $H(\mathbf{k})$ that depends on wave vector $\mathbf{k}$ in the Brillouin zone \cite{newdiracpoints}: \begin{equation}
        H = \begin{pmatrix}
                \Delta_1 + \Delta_2 & \gamma_0 t^*_k                & \gamma_4 t^*_k    & \gamma_3 t_k      & \frac{\gamma_2}{2}    & 0 \\
                \gamma_0 t_k        & \delta + \Delta_1 + \Delta_2  & \gamma_1          & \gamma_4 t^*_k    & 0                     & \frac{\gamma_5}{2} \\
                \gamma_4 t_k        & \gamma_1                      & \delta -2\Delta_2 & \gamma_0 t^*_k    & \gamma_4 t_k          & \gamma_1 \\
                \gamma_3 t^*_k      & \gamma_4 t_k                  & \gamma_0 t_k      & -2\Delta_2        & \gamma_3 t^*_k        & \gamma_4 t_k \\
                \frac{\gamma_2}{2}  & 0                             & \gamma_4 t^*_k    & \gamma_3 t_k      & -\Delta_1 + \Delta_2  & \gamma_0 t^*_k \\
                0                   & \frac{\gamma_5}{2}            & \gamma_1          & \gamma_4 t^*_k    & \gamma_0 t_k          & \delta -\Delta_1 +\Delta_2 \\
            \end{pmatrix}
\end{equation}

A multitude of parameters show up in $H$, which we now explain. We have defined $t_k$ to be 
\begin{equation}\label{eq:tk}
    t_k = e^{i \vec{k} \cdot \vec{\delta}_1} + e^{i \vec{k} \cdot \vec{\delta}_2} + e^{i \vec{k} \cdot \vec{\delta}_3} = -1 - 2e^{\frac{\sqrt{3}}{2}ik_y} \cos{\frac{k_x}{2}}
\end{equation}
where $\vec{\delta}_1 = \frac{a}{\sqrt{3}}(0,1)$, $\vec{\delta}_{2,3} = \frac{a}{\sqrt{3}} (\mp \sqrt{3}/2, -1/2)$. Here $a \approx 0.246$ nm is the lattice constant.
There are $6$ hopping parameters $\gamma_0, \ldots \gamma_5$, which correspond to hopping matrix elements between differing sites both within a layer and between layers (interlayer hoppings are shown in Figure \ref{fig:aba}):
\begin{subequations}
\begin{align}
        A_i \leftrightarrow B_i: &\gamma_0 \\
        A_2 \leftrightarrow B_{1,3}: &\gamma_1 \\
        A_1 \leftrightarrow A_3: &\frac{1}{2} \gamma_2 \\
        A_{1,3} \leftrightarrow B_2:  &\gamma_3 \\
        A_{1,3} \leftrightarrow A_2: -&\gamma_4 \\
        B_{1,3} \leftrightarrow B_2: -&\gamma_4 \\
        B_1     \leftrightarrow B_3: &\frac{1}{2} \gamma_5 
\end{align}
\label{eq:hopping}
\end{subequations}
Next, there is an on-site energy $\delta$ which arises due to the overlap in orbitals on the $B_1$, $A_2$, and $B_3$ sites. Finally, the effects of perpendicular electric fields which come from electronic gates are encoded in potentials $\Delta_1$ and $\Delta_2$. In terms of layer potentials $U_{1,2,3}$, 
\begin{subequations}
\begin{align}
        \Delta_1 &= (-e) \frac{U_1-U_3}{2} \\
        \Delta_2 &= (-e) \frac{U_1 - 2U_2 + U_3}{6}
\end{align}
\end{subequations}

In the absence of $\Delta_1$, there is a layer exchange symmetry $A_1 \leftrightarrow A_3$, $B_1 \leftrightarrow B_3$ which interchanges the outer layers. In the literature, it is common to rewrite the tight-binding Hamiltonian in a basis that diagonalizes that layer exchange symmetry. We will moreover be interested in the physics near the following two points in the Brillouin zone: \begin{equation}
    K^\pm = \pm \frac{4\pi}{3a} (0,1).
\end{equation}In the $\{\frac{A_1-A_3}{\sqrt{2}}, \frac{B_1-B_3}{\sqrt{2}}, \frac{A_1+A_3}{\sqrt{2}}, B_2, A_2, \frac{B_1+B_3}{\sqrt{2}}\}$ basis, the Hamiltonian expanded around the $K^\pm$ point is  
\renewcommand{\dag}[1]{#1^\dagger}
\begin{equation}
        H_0 =   \begin{pmatrix}
                \Delta_2 - \gamma_2/2 & v_0 \dag{\pi} & \Delta_1 & 0 & 0 & 0 \\
                v_0 \pi & \delta + \Delta_2 - \gamma_5/2 & 0 & 0 & 0& \Delta_1 \\
                \Delta_1 & 0 & \Delta_2 + \gamma_2 & \sqrt{2} v_3 \pi & -\sqrt{2} v_4 \dag{\pi} & v_0 \dag{\pi} \\
                0 & 0 & \sqrt{2} v_3 \dag{\pi} & -2 \Delta_2 & v_0 \pi & -\sqrt{2} v_4 \pi \\
                0 & 0 & -\sqrt{2} v_4 \pi & v_0 \dag{\pi} & \delta - 2\Delta_2 & \sqrt{2} \gamma_1 \\
                0 & \Delta_1 & v_0\pi & -\sqrt{2} v_4 \dag{\pi} & \sqrt{2} \gamma_1 & \delta + \Delta_2 + \gamma_5/2 
                \end{pmatrix}
\end{equation}
The velocities $v_i$ are defined as $v_i = \sqrt{3}a\gamma_i/(2\hbar)$ , while $\pi = p_x \pm \mathrm{i}p_y$ is the deviation from the $K^{\pm}$ point.

\begin{figure}
        \centering
        \begin{subfigure}[b]{0.32\textwidth}
            \caption{$E_F = 0.005$ eV}
            \includegraphics[width=\textwidth]{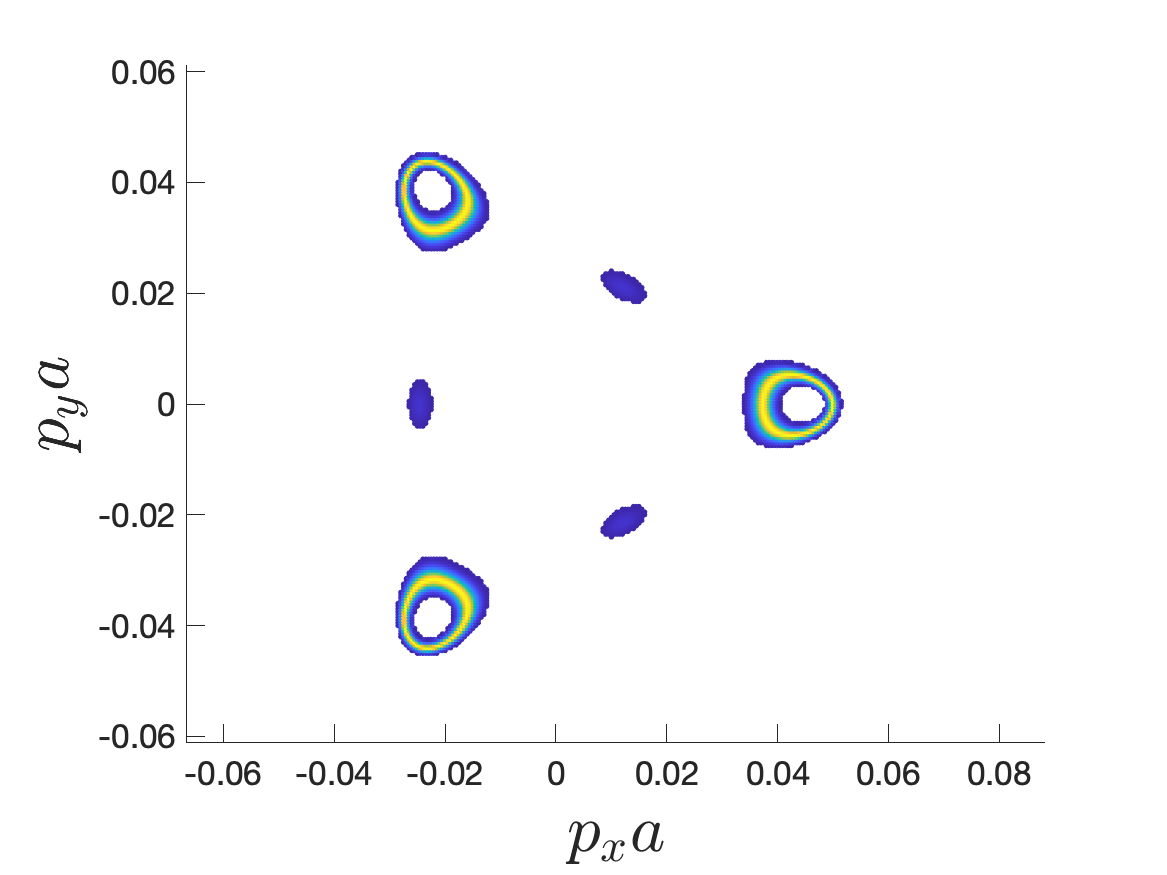}
            \label{fig:disconnectedFS}
        \end{subfigure}
        \begin{subfigure}[b]{0.32\textwidth}
            \caption{$E_F = 0.01$ eV}
            \includegraphics[width=\textwidth]{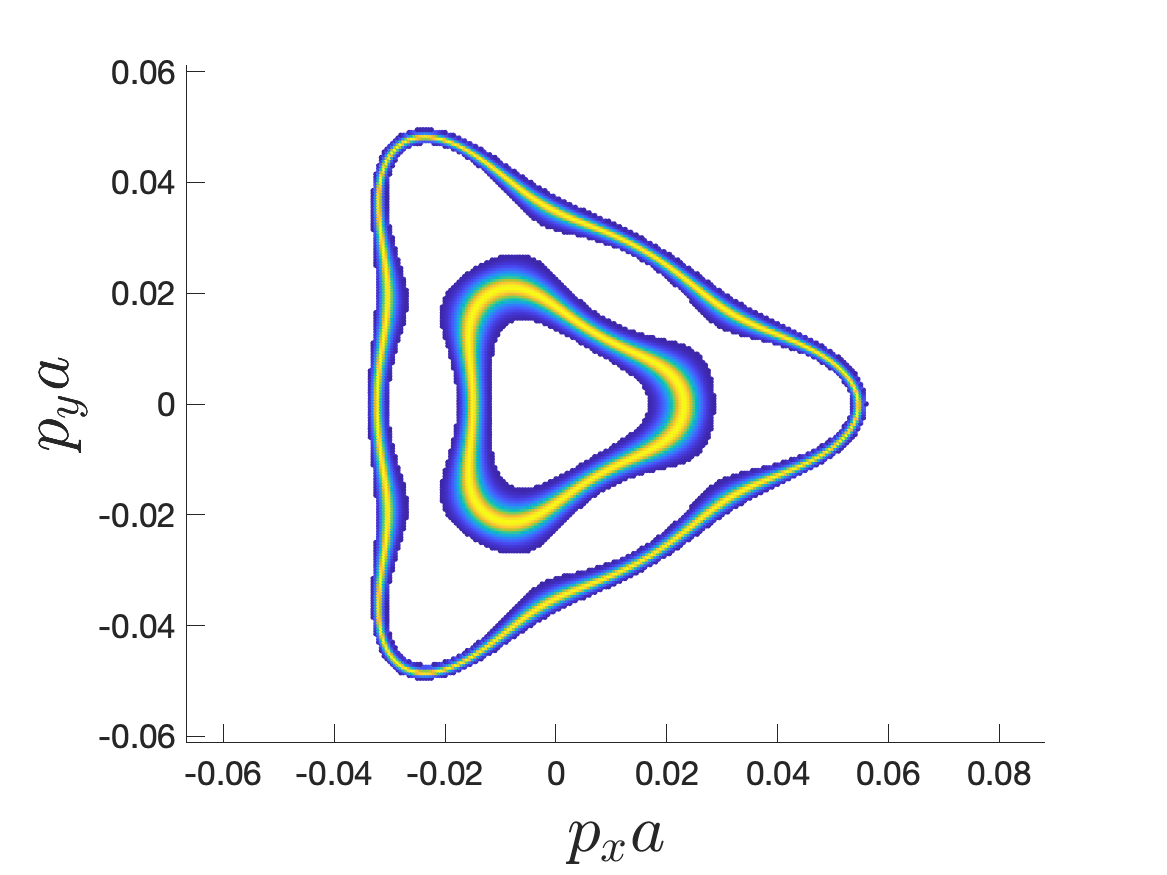}
            \label{fig:pocketFS}
        \end{subfigure}
        \begin{subfigure}[b]{0.32\textwidth}
            \caption{$E_F = 0.02$ eV}            \includegraphics[width=\textwidth]{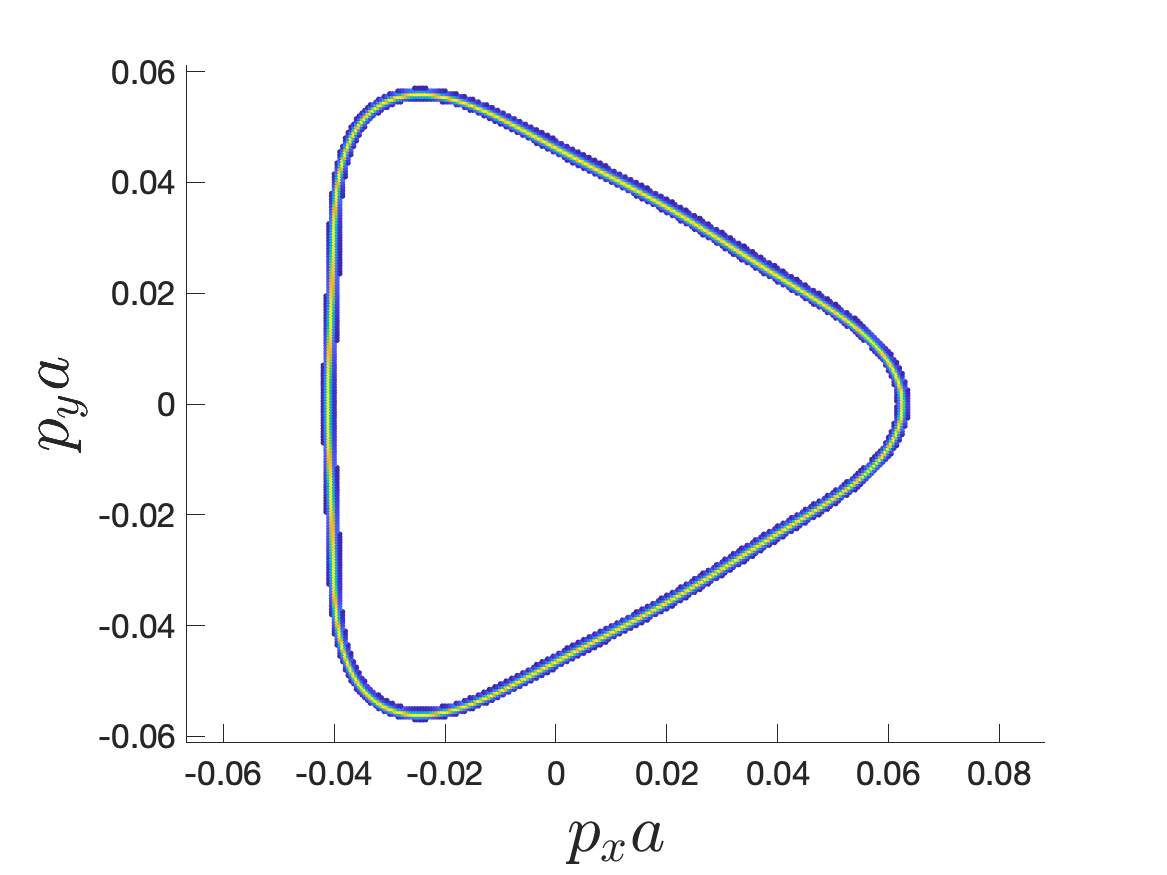}
            \label{fig:triangleFS}
        \end{subfigure}
        \caption{Thermally smeared Fermi surfaces near the $K^+$ point. Color denotes proximity to the Fermi surface, with yellow denoting states close to the Fermi surface and blue denoting thermally excited states further away. The Fermi energies are $0.005 \text{ eV}$ (left), $0.01 \text{ eV}$ (middle), and $0.02 \text{ eV}$ (right). }
        \label{fig:fermisurfaces}
\end{figure}

For the parameters $\gamma_i$, $\delta$, and $\Delta_2$, we use the values found in Table S1 of \cite{abatrilayer}, and we take $\Delta_1 = 80 \text{meV}$. The resulting band structure is shown in Figure \ref{fig:fermisurfaces}. As we tune the Fermi energy, we find interesting Fermi surface configurations, as shown in Figure \ref{fig:fermisurfaces}; these of course correspond to the level surfaces of the six-band dispersion relation $\epsilon_a(k_x,k_y)$ ($a=1,\ldots, 6$).   Similar Fermi surfaces were found in \cite{abatrilayer} by tuning $\Delta_1$ and the charge density.   Since for the parameters chosen, only one band has $\epsilon_a(k_x,k_y) = E_{\mathrm{F}}$ at any given $(k_x,k_y)$, the dynamics of electrons close to the Fermi level is essentially captured by a single-band model with non-trivial Fermi surface geometry and/or topology. We emphasize that the latter two Fermi surfaces of Figure \ref{fig:fermisurfaces} are approximately polygonal, with a large fraction of the Fermi surface having parallel Fermi velocity: the consequences of this for ballistic transport will become clear later.

An important observation from Figure \ref{fig:fermisurfaces} is that in all three Fermi surfaces we will study in this paper, the values of $p_x a$ are very small compared to 1 (the scale at which umklapp processes would become non-negligible).  As a consequence, we do not expect umklapp to be more significant in this system than it would have been in monolayer or bilayer graphene.

\subsection{Numerical solutions of the kinetic equations}
\label{sec:abaflow}

Following the framework of Section \ref{sec:kinetictheory}, we numerically compute the Green's function $\mathsf{G}(\mathbf{k})$ for the band structure describing ABA graphene. Using the corresponding conductivity $\sigma_{ij}(\mathbf{k})$, we obtain the current profile $\langle \vec{J}(x)\rangle$ for the constriction geometry of Fig. \ref{fig:constriction}, following the methods of Section \ref{sec:method}. To probe the different transport regimes, we study the resulting flow patterns for different electron-electron and electron-impurity scattering lengths. 

As discussed in the introduction, it is expected that there are three distinct transport regimes in this system, which are controlled by the relative sizes of the electron-electron interaction length $\ell_{\mathrm{ee}}$, the electron-impurity interaction length $\ell_{\mathrm{ei}}$, and the constriction width $w$. In particular, diffusive (ohmic) flows should occur when $\sqrt{\ell_{\mathrm{ei}}\ell_{\mathrm{ee}}} \ll w$, viscous flows occur when $\ell_{\mathrm{ee}} \ll w \lesssim \sqrt{\ell_{\mathrm{ei}}\ell_{\mathrm{ee}}}$, and ballistic flows occur when $w \lesssim \ell_{\mathrm{ee}}, \ell_{\mathrm{ei}}$. Here, we show that this expectation is indeed satisfied (up to some subtleties about O(1) prefactors which we discuss in the next section).  Moreover, we will see that the different transport regimes can be identified from qualitative features of $\langle \vec{J}(x) \rangle$. In particular, the anisotropy of the Fermi surface allows us to sharply distinguish between the viscous and ballistic flows.

\begin{figure}
    \centering
    \begin{subfigure}{0.32\textwidth}
        \subcaption{$\lee = 0.05w$, $\lei = 0.2w$}
        \vspace{-7pt}
        \includegraphics[width=\linewidth]{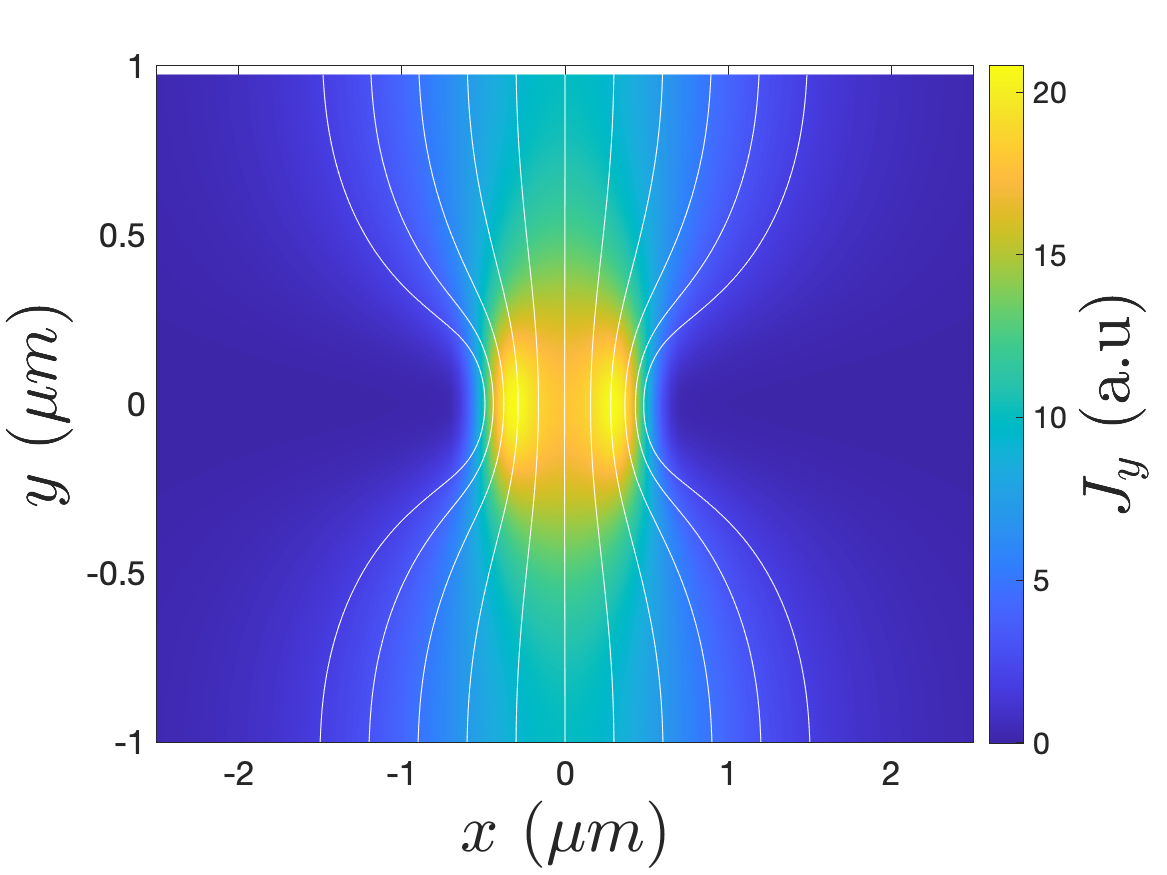}
        \label{fig:disconnected_ohmic}
    \end{subfigure}
    \begin{subfigure}{0.32\textwidth}
        \subcaption{$\lee = 0.05w$, $\lei = 2w$}
        \vspace{-7pt}
        \includegraphics[width=\linewidth]{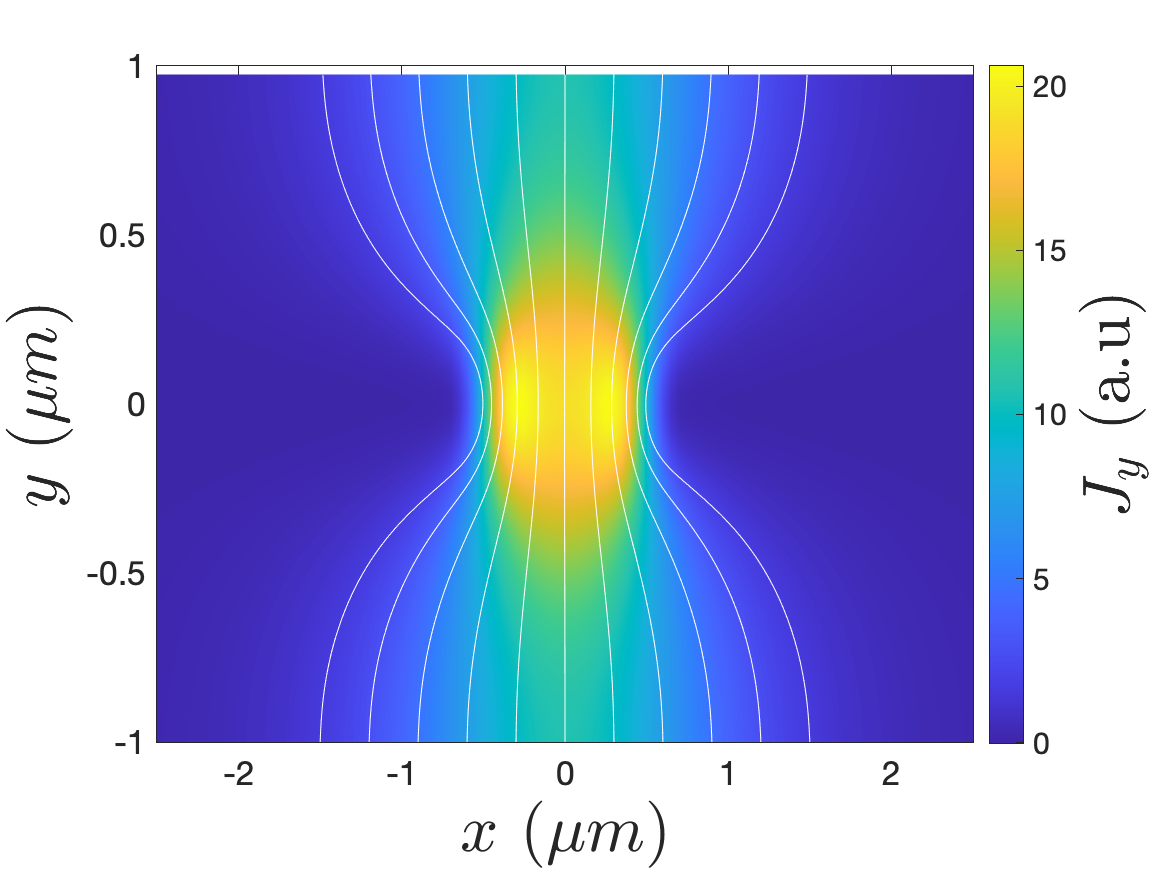}
        \label{fig:disconnected_hydro}
    \end{subfigure}
    \begin{subfigure}{0.32\textwidth}
        \subcaption{$\lee = w$, $\lei = 2w$}
        \vspace{-7pt}
        \includegraphics[width=\linewidth]{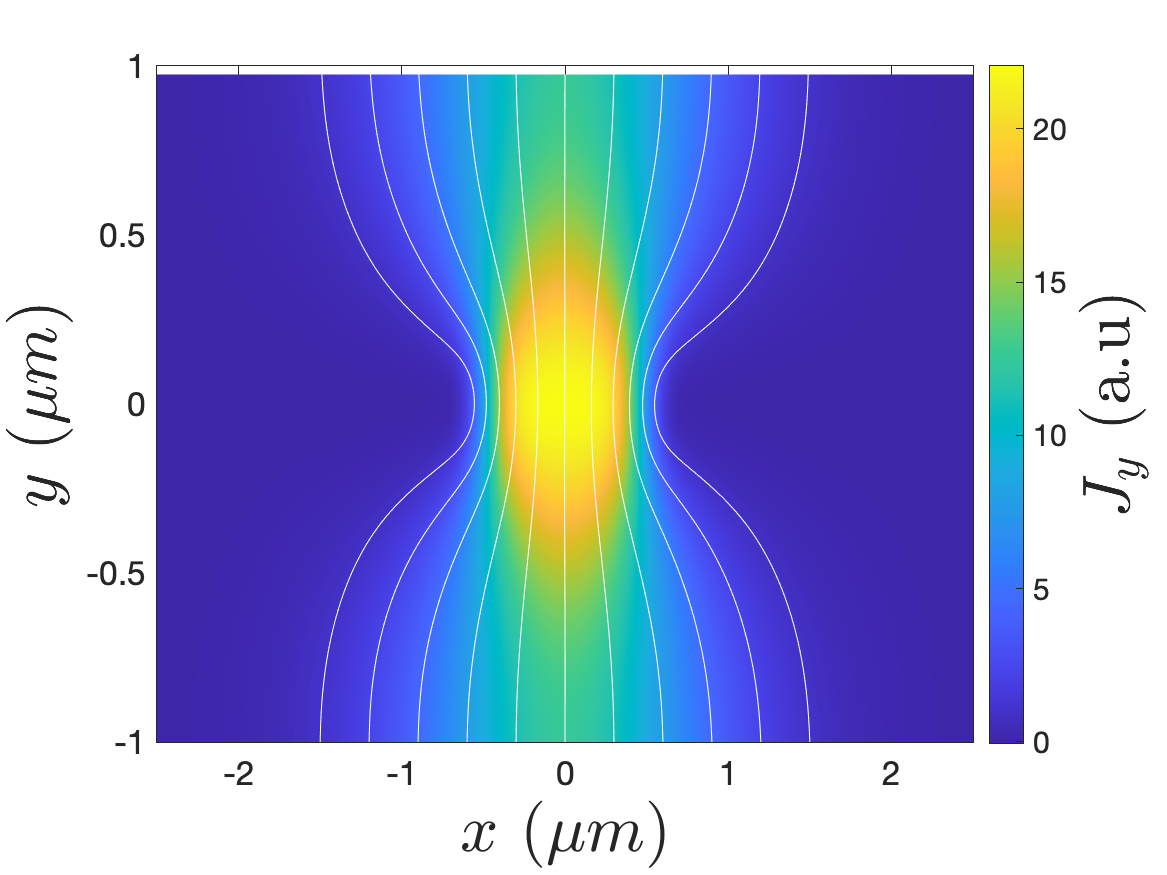}
        \label{fig:disconnected_ballistic}
    \end{subfigure}
    \caption{Flow patterns with $\lee = 0.05w$, $\lei = 0.2w$ (left), $\lee = 0.05w$, $\lei = 2w$ (middle), and $\lee = w$, $\lei = 2w$ (right) for the Fermi surface \ref{fig:disconnectedFS}.}
    \label{fig:flow_disconnected}
\end{figure}
\begin{figure}
    \centering
    \begin{subfigure}{0.32\textwidth}
        \subcaption{$\lee = 0.05w$, $\lei = 0.2w$}
        \vspace{-7pt}
        \includegraphics[width=\linewidth]{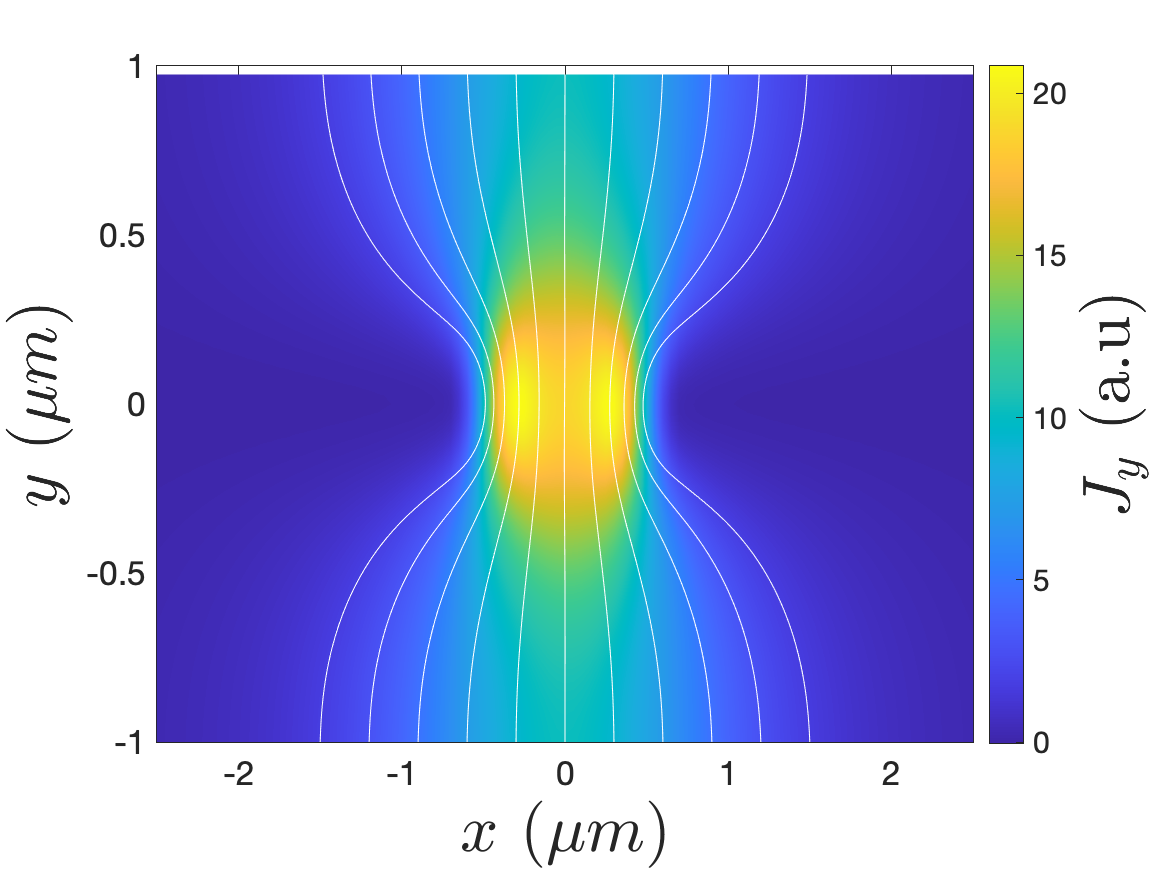}
        \label{fig:pocket_ohmic}
    \end{subfigure}
    \begin{subfigure}{0.32\textwidth}
        \subcaption{$\lee = 0.05w$, $\lei = 2w$}
        \vspace{-7pt}
        \includegraphics[width=\linewidth]{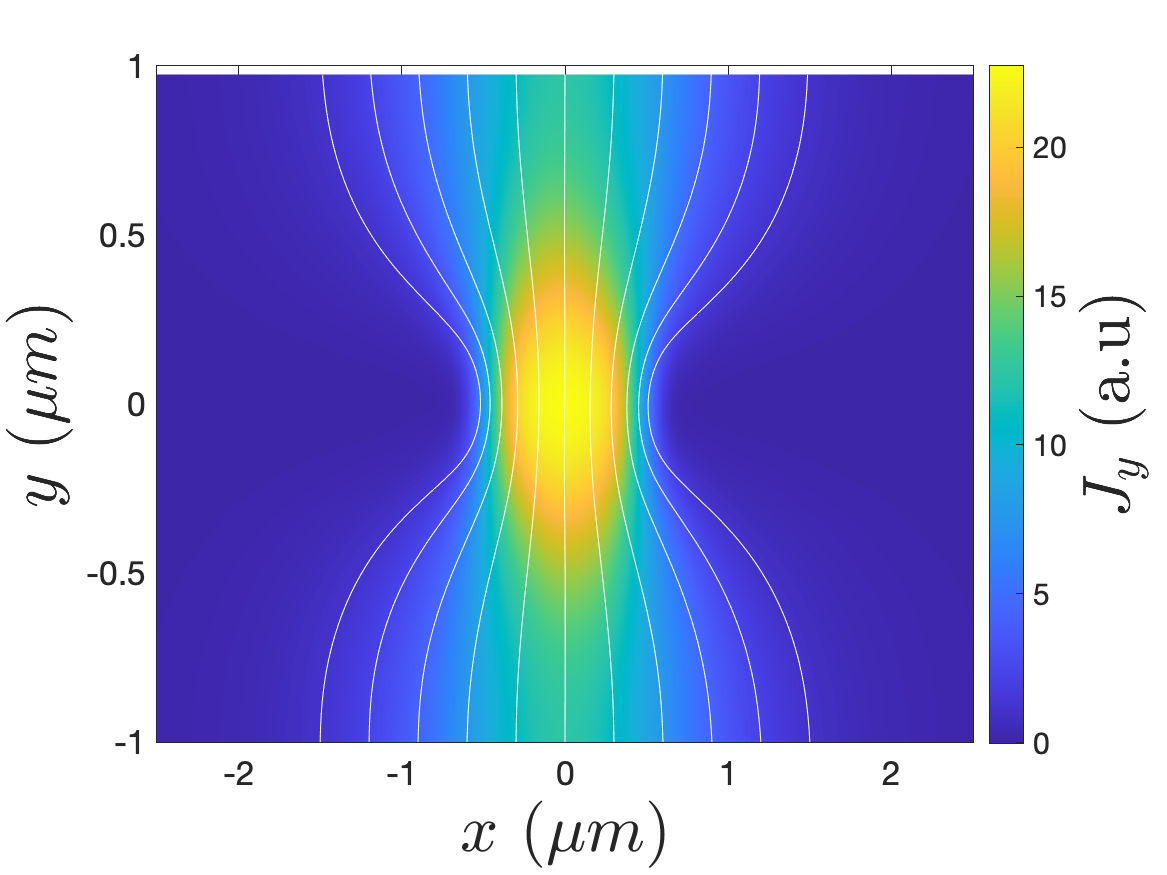}
        \label{fig:pocket_hydro}
    \end{subfigure}
    \begin{subfigure}{0.32\textwidth}
        \subcaption{$\lee = w$, $\lei = 2w$}
        \vspace{-7pt}
        \includegraphics[width=\linewidth]{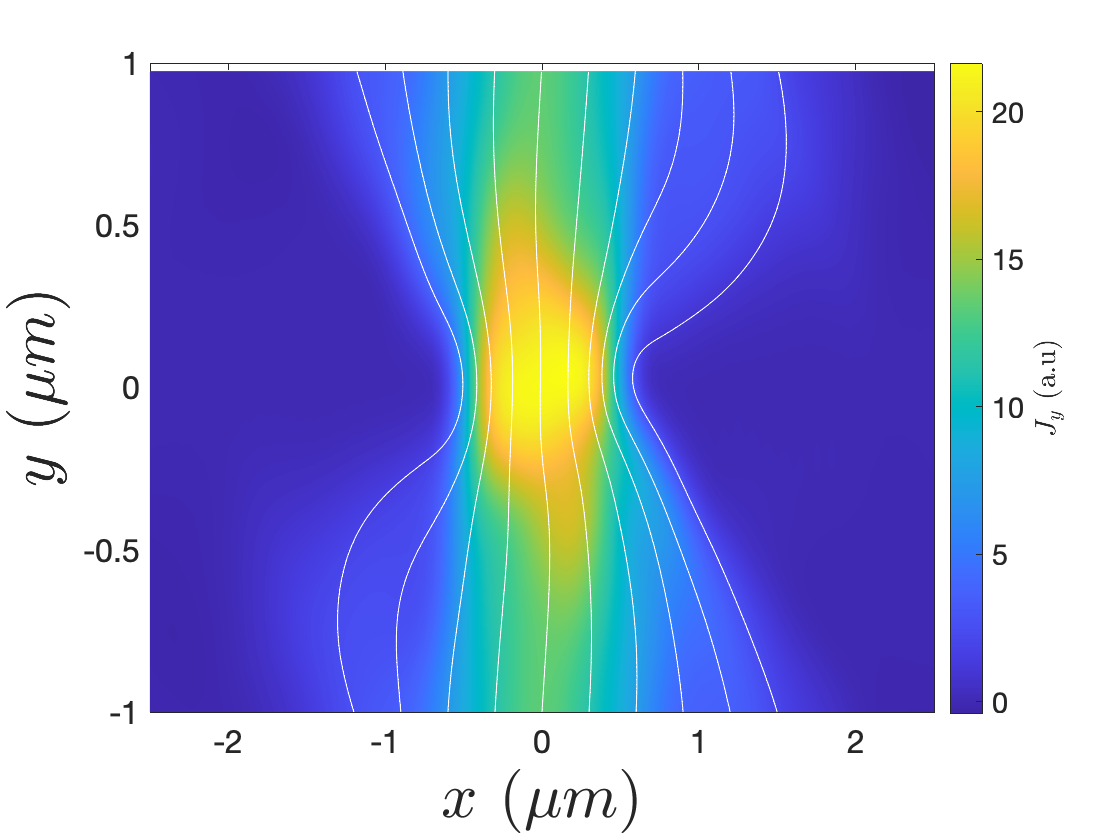}
        \label{fig:pocket_ballistic}
    \end{subfigure}
    \caption{Flow patterns with $\lee = 0.05w$, $\lei = 0.2w$ (left), $\lee = 0.05w$, $\lei = 2w$ (middle), and $\lee = w$, $\lei = 2w$ (right) for the Fermi surface \ref{fig:pocketFS}.}
    \label{fig:flow_pocket}
\end{figure}
\begin{figure}[t]
    \centering
    \begin{subfigure}{0.32\textwidth}
        \subcaption{$\lee = 0.05w$, $\lei = 0.2w$}
        \vspace{-7pt}
        \includegraphics[width=\linewidth]{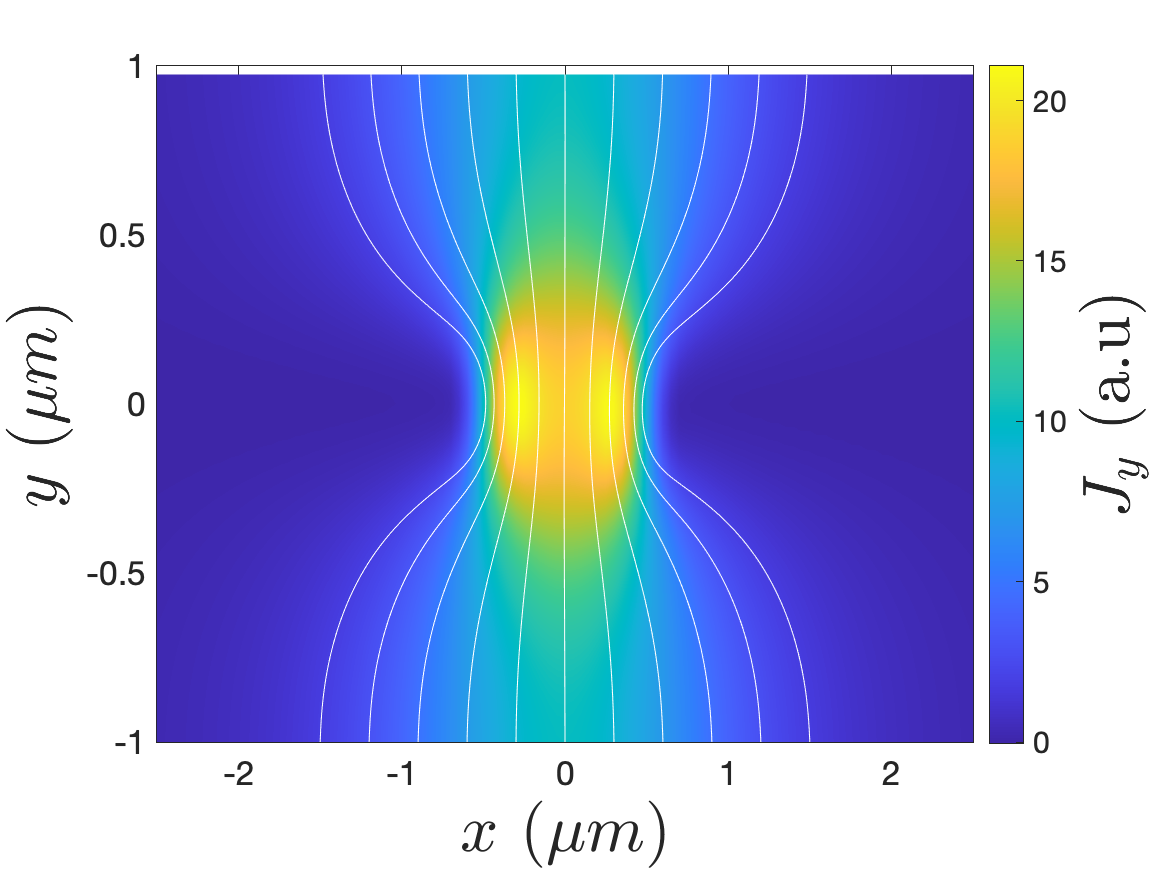}
        \label{fig:triangle_ohmic}
    \end{subfigure}
    \begin{subfigure}{0.32\textwidth}
        \subcaption{$\lee = 0.05w$, $\lei = 2w$}
        \vspace{-7pt}
        \includegraphics[width=\linewidth]{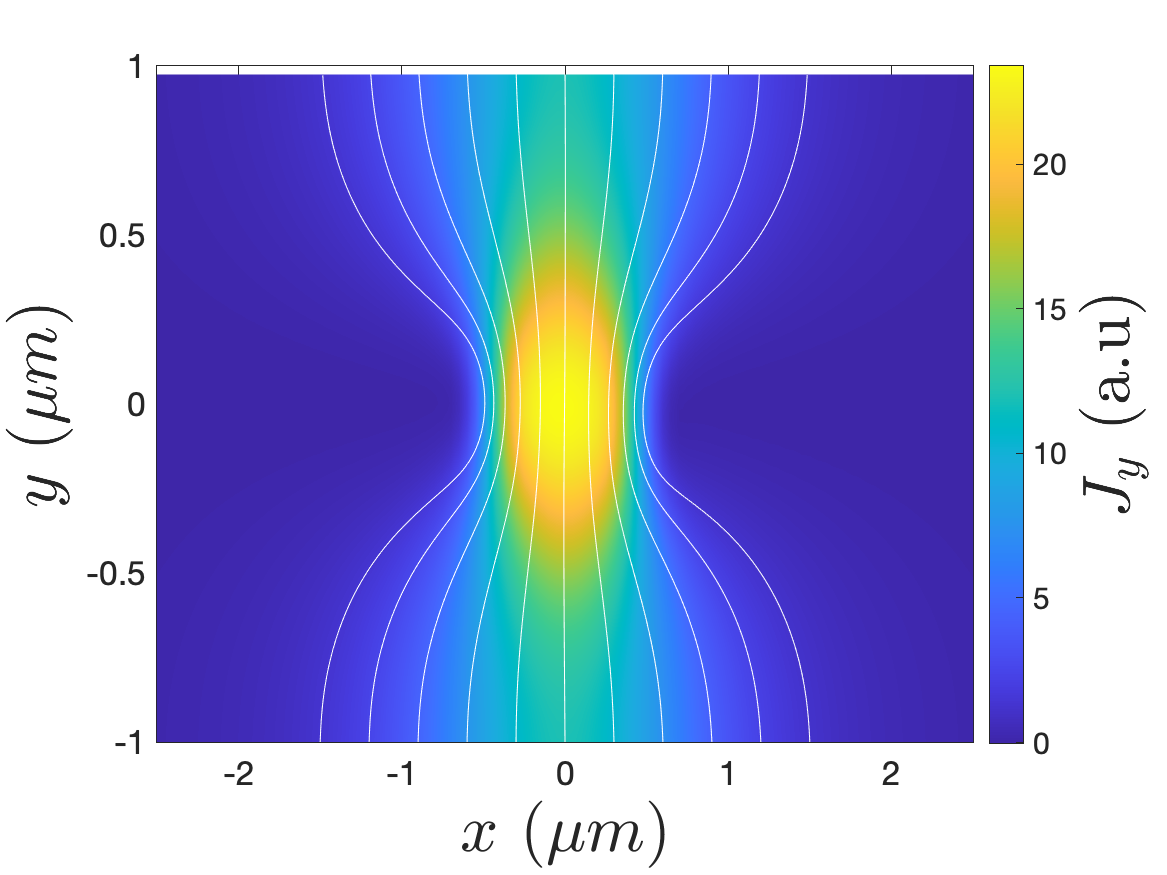}
        \label{fig:triangle_hydro}
    \end{subfigure}
    \begin{subfigure}{0.32\textwidth}
        \subcaption{$\lee = w$, $\lei = 2w$}
        \vspace{-7pt}
        \includegraphics[width=\linewidth]{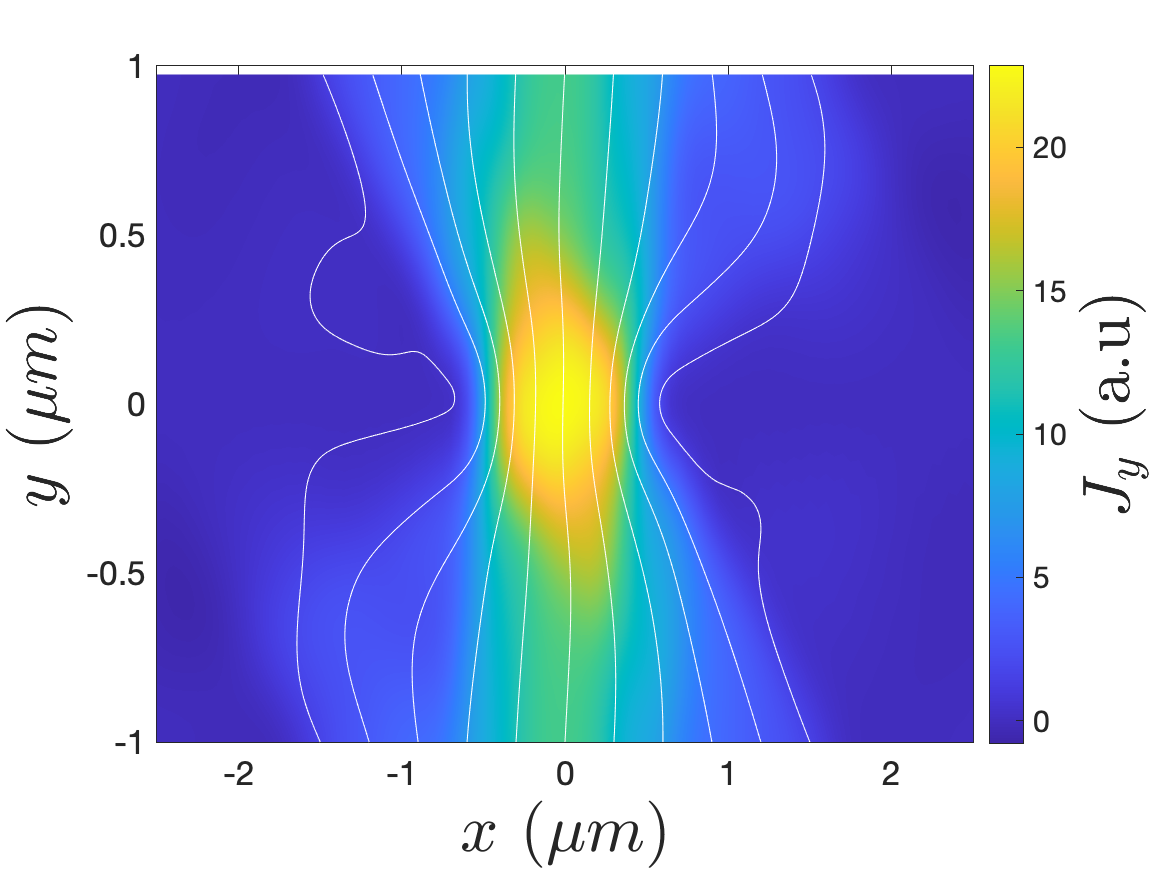}
        \label{fig:triangle:ballistic}
    \end{subfigure}
    \caption{Flow patterns with $\lee = 0.05w$, $\lei = 0.2w$ (left), $\lee = 0.05w$, $\lei = 2w$ (middle), and $\lee = w$, $\lei = 2w$ (right) for the Fermi surface \ref{fig:triangleFS}.}
    \label{fig:flow_triangle}
\end{figure}

The flow patterns for the Fermi surfaces of Figure \ref{fig:fermisurfaces} for various $\lee$ and $\lei$ are shown in Figures \ref{fig:flow_disconnected}-\ref{fig:flow_triangle}. Figures \ref{fig:flow_disconnected}, \ref{fig:flow_pocket}, and \ref{fig:flow_triangle} correspond to the Fermi surfaces shown in Figures \ref{fig:disconnectedFS}, \ref{fig:pocketFS}, and \ref{fig:triangleFS} respectively. Within each figure, the length scales are varied so that the system is expected to be in the ohmic, viscous, and ballistic regimes in the first (a), second (b), and third (c) subfigures respectively.  For each geometry, we calculate $\lee$ and $\lei$ as follows: \begin{equation}
        \sqrt{\langle v(p)^2\rangle_{\mathrm{FS}}} = \lee \gamma_{\mathrm{ee}} = \lei \gamma_{\mathrm{ei}} 
        \end{equation} 
        The average of the quasiparticle velocity squared near the Fermi surface is numerically calculated by averaging over points within an energy $T_* = 3$K of the Fermi energy $E_{\mathrm{F}}$:  $T_* \gtrsim |\epsilon(p)-E_{\mathrm{F}}|$, as described at the end of Section \ref{sec:kinetictheory}.  We chose this extremely small value of $T_*$ because it renders the Fermi surface sharp.  While in principle our approach can be generalized to Fermi liquids with $T\sim T_{\mathrm{F}}$, one must also more carefully account for energy conservation (which is subleading in $T/T_{\mathrm{F}}$ in our current calculation), \emph{and} the computational costs are substantially higher (as one can no longer neglect points in most of the Brillouin zone when numerically evaluating integrals over it).   Ultimately, this $T_*$ plays the role of an effective temperature, although quantitatively this is not entirely appropriate (we will return to this point at the end of Section \ref{sec:compare}).
        In calculating $\mathsf{G}(\mathbf{k})$, the Fermi surfaces were rotated $0.35$ radians relative to the constriction.  This rotation is important to easily identify the ballistic regime, and as we will see, for imaging ballistic flows it is best to \emph{not} align the crystal lattice in the material with the device geometry.

Figures \ref{fig:disconnected_ohmic}, \ref{fig:pocket_ohmic}, and \ref{fig:triangle_ohmic} show the flow patterns for $\lee = 0.05w$, $\lei = 0.2w$, where the flow behaves diffusively for all three Fermi surfaces. As in \cite{jenkins2020imaging}, the flow is characterized by peaks in the current density near the edges of the constriction.  This result is universal and follows from the solutions of the Laplace equation around sharp corners (the ``lightning rod effect") \cite{jenkins2020imaging}. The anisotropy of the Fermi surface and the relative orientation do not strongly affect the flow pattern in this limit.

Figures \ref{fig:disconnected_hydro}, \ref{fig:pocket_hydro}, and \ref{fig:triangle_hydro} show the flow patterns for $\lee = 0.05w, \lei = 2w$. The length scales suggest that the flow should be weakly viscous; however, this is true only for the latter two Fermi surfaces. The flow for the first Fermi surface is more similar to the diffusive flows in the first row. For the latter two Fermi surfaces, there is a single peak in the current density at the center of the constriction, sharply distinguishing it from the diffusive flows in the first row. The flows remain approximately reflection symmetric across the center of the constriction, despite the relative orientation between the Fermi surface and constriction breaking the symmetry microscopically -- we will return to this point later.

Figures \ref{fig:disconnected_ballistic}, \ref{fig:pocket_ballistic}, and \ref{fig:triangle:ballistic} show the flow patterns for $\lee = w$, $\lei = 2w$. For all three Fermi surfaces, the flow exhibits angular streaks of high current density suggestive of ballistic transport. Furthermore, for the second and third Fermi surfaces, the streaks are tilted relative to the constriction; they can also be seen via the tilt of the streamlines. Together, they provide a clear signal for ballistic transport. For the disconnected Fermi surface, the asymmetry of the streaks is suppressed because each island of the Fermi surface is approximately isotropic. The approximately polygonal Fermi surface and its orientation relative to the constriction is crucial for the phenomena observed in Figures \ref{fig:pocket_ballistic} and \ref{fig:triangle:ballistic}, as we will discuss later.

\begin{center}
    \begin{figure}
        \centering
        \begin{tabular}{c c c}
        \includegraphics[width=0.33\textwidth]{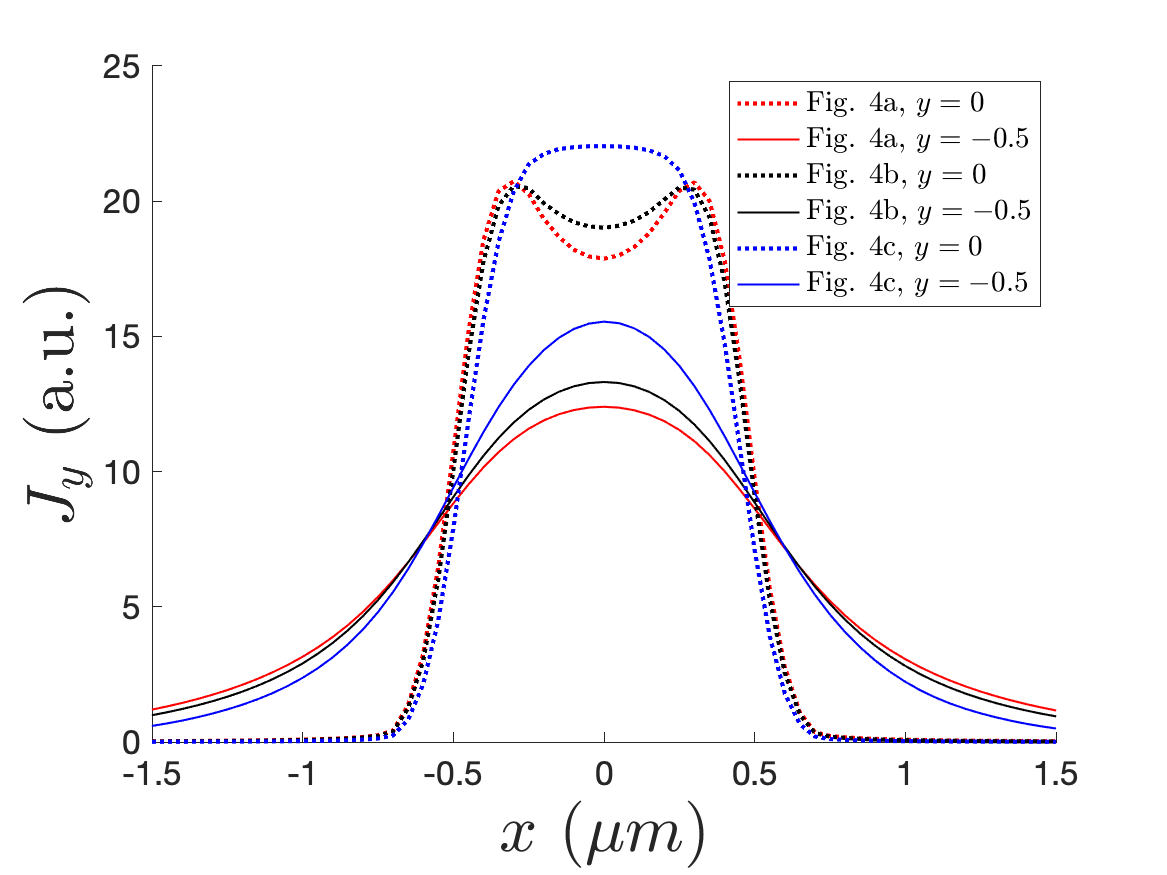} &
        \includegraphics[width=0.33\textwidth]{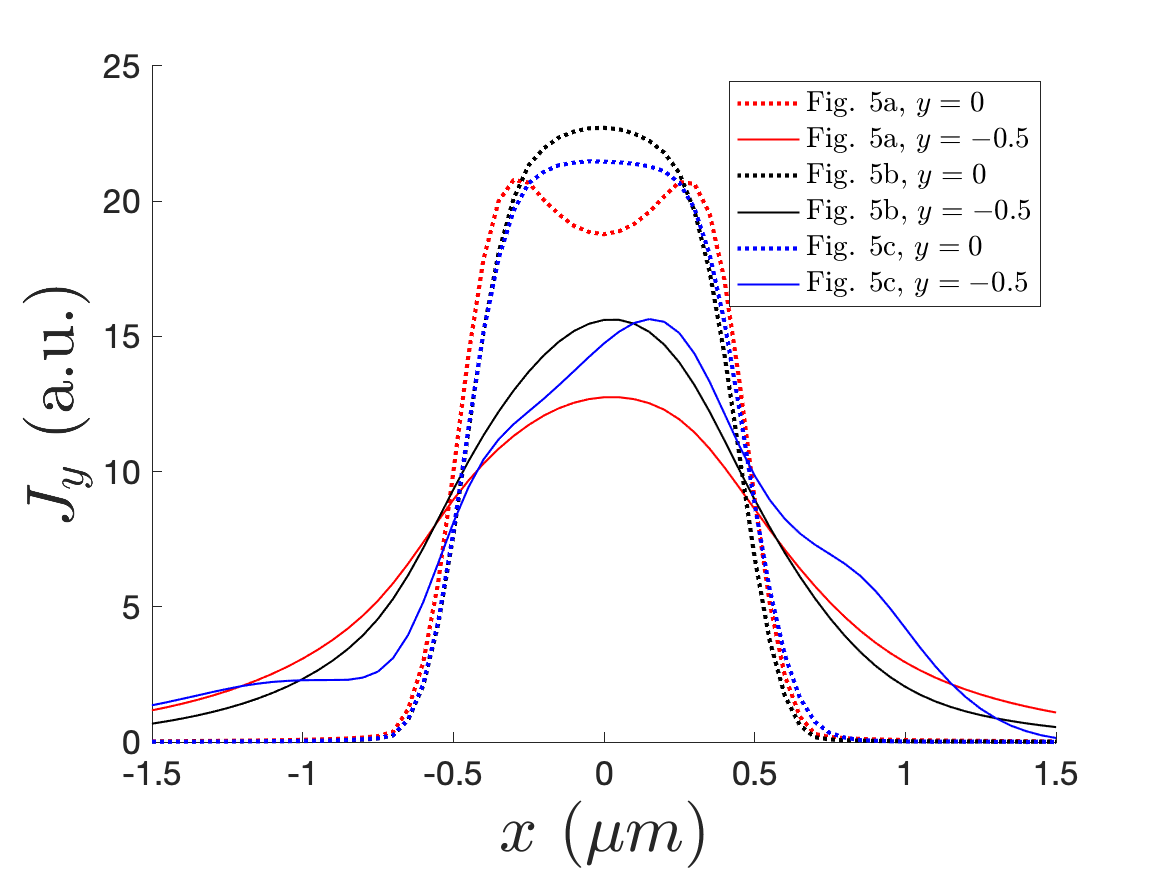} &
        \includegraphics[width=0.33\textwidth]{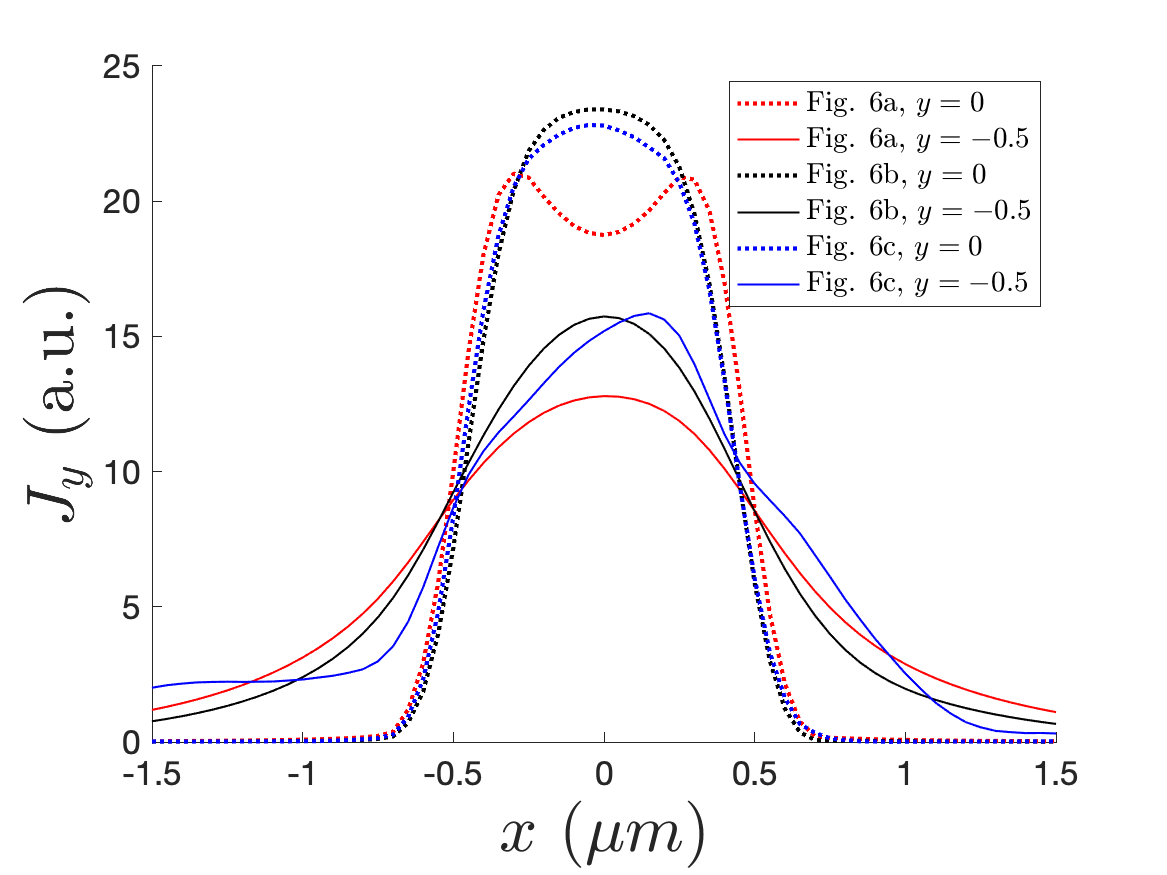}
        \end{tabular}
        \caption{Flow profiles $J_y$ vs. $x$ through the center of the slit ($y=0$) and through $y=-0.5 \; \mu\mathrm{m}$ for the Fermi surfaces in Figure \ref{fig:fermisurfaces}. The left, middle, and right plots correspond to Fermi surfaces \ref{fig:disconnectedFS}, \ref{fig:pocketFS}, and \ref{fig:triangleFS}, and correspond to the data shown in Figures \ref{fig:flow_disconnected}, \ref{fig:flow_pocket}, and \ref{fig:flow_triangle}, respectively. }
        \label{fig:currentxc}
    \end{figure}
\end{center}

To the human eye, it can be somewhat challenging to detect the differences between the various flow patterns without knowing exactly what to look for.  Thus, as a ``guide to the eye", we have also plotted a few cross-sections of flow patterns along lines of fixed $y$-coordinate, where we can easily discern patterns that are unique to each of our three transport regimes.  For example, distinguishing the viscous and ballistic regimes can be done by looking at one dimensional cross sections of the flow plots through $y=0$ and $y=\pm 0.5 \; \mu \mathrm{m}$; distinguishing ohmic and viscous regimes is best done by studying flows through $y=0$, where there will be two maxima in the current profile in the ohmic regime, versus one in the viscous regime. Figure \ref{fig:currentxc} plots $J_y$ vs. $x$ for $y=0$ and $y=-0.5\; \mu \mathrm{m}$ for the the different transport regimes and Fermi surfaces of Figure \ref{fig:fermisurfaces}. For the latter two Fermi surfaces, the ballistic regime can be identified by the offset of the peak from $x=0$ for the current cross section through $y=-0.5 \; \mu \mathrm{m}$.

\newcommand{\putpic}[1]{\includegraphics[width=0.33\textwidth]{#1}}

\subsection{Hydrodynamics of a hexagonal electron fluid}
To interpret the results which we have found above, we now turn to a discussion of the nature of hydrodynamics in an electron fluid with the symmetry of ABA graphene (the dihedral group $\mathrm{D}_{6}$ in physics literature, and $\mathrm{D}_{12}$ in mathematical literature).  This anisotropic hydrodynamics was first derived in \cite{caleb}.  The linearized Navier-Stokes equations take the form \begin{subequations}
    \begin{align}
        \partial_t n + \partial_i \left(n_0 v_i - \sigma_0 \partial_i \mu \right) &= 0, \\
        \partial_t g_i + \partial_j \left( m c_{\mathrm{s}}^2 n\delta_{ij} - \eta_{jikl}\partial_k v_l\right) &= 0,
    \end{align}
\end{subequations}
where $n$ denotes the hydrodynamic number density of electrons (with $n_0\ne 0$ the equilibrium value, as measured relative to $E_{\mathrm{F}}=0$), $m$ denotes an ``averaged" quasiparticle mass, $g_i$ the momentum  density, $\sigma_0$ an ``incoherent conductivity", $c_{\mathrm{s}}$ the speed of (electronic) sound, and $\eta_{jikl}$ the viscosity tensor.

The incoherent conductivity $\sigma_0$ arises due to the inequivalence of the charge current $J$ and the momentum density $g_i$:  in the kinetic theory formalism, $|\mathsf{J}_i\rangle \ne |\mathsf{P}_i\rangle$.   $\sigma_0$ thus corresponds to a decay mechanism for density imbalance that arises through the relaxation of the non-conserved part of the current operator which is orthogonal to momentum.    The viscosity tensor $\eta_{jikl}$ encodes the dissipation of momentum, and takes the following form \begin{equation}
    \eta_{jikl}  = \eta \left(\delta_{ik}\delta_{jl} + \delta_{il}\delta_{jk} - \delta_{ij}\delta_{kl}\right) + \zeta \delta_{ij}\delta_{kl} + \tilde\eta \epsilon_{ij}\epsilon_{kl}.
\end{equation}
Here $\delta_{ij}$ is the Kronecker $\delta$ function and $\epsilon_{ij}$ is the two-dimensional Levi-Civita tensor, while $\eta$, $\zeta$ and $\tilde\eta$ denote the shear, bulk, and rotational viscosities respectively.  The rotational viscosity arises due to the explicitly broken continuous rotational symmetry (due to the anisotropy of the dispersion relation and Fermi surface): for a detailed discussion of this term, see \cite{caleb}.

Independently of our interest in imaging transport, we may ask what these dissipative hydrodynamic coefficients even are in our relaxation time model of ABA graphene.  These coefficients may be straightforwardly calculated in the kinetic formalism of Section \ref{sec:kinetictheory}: defining the vectors 
\begin{subequations}\begin{align}
|\mathsf{J}_i\rangle &= \int\mathrm{d}^2p \; v_i |p\rangle , \\
|\mathsf{T}_{ij}\rangle &= \int\mathrm{d}^2p \;  v_i p_j |p\rangle ,
\end{align}\end{subequations}
we find that 
\begin{subequations}\begin{align}
    \eta_{ijkl} &= \langle \mathsf{T}_{ij} | (1-\mathsf{P}) \mathsf{W}^{-1} (1-\mathsf{P}) | \mathsf{T}_{kl} \rangle , \\
    \sigma_0 \delta_{ij} &= \langle \mathsf{J}_i | (1-\mathsf{P})  \mathsf{W}^{-1} (1-\mathsf{P})  | \mathsf{J}_j \rangle.
\end{align}\end{subequations}
The projector $\mathsf{P}$ was defined in (\ref{eq:projector}).  These equations are in fact quite general, and do not rely on any relaxation time approximation: using the relaxation time approximation, we see that \begin{subequations}\begin{align}
    \eta_{ijkl} &= \frac{1}{\gamma_{\mathrm{ee}}} \langle \mathsf{T}_{ij} | 1 - \mathsf{P} | \mathsf{T}_{kl} \rangle \\
    \sigma_{ij} &= \frac{1}{\gamma_{\mathrm{ee}}}  \langle \mathsf{J}_i | 1 - \mathsf{P} | \mathsf{J}_j \rangle.
\end{align} \end{subequations}

We have numerically calculated the four numbers $\sigma_0$, $\eta$, $\zeta$ and $\tilde\eta$ for each of the three Fermi surfaces shown above.  In doing this calculation we have indeed confirmed that the conductivity and viscosity tensors have the symmetry predicted above, with only 4 total distinct coefficients.  In order to compare between Fermi surfaces, we normalize the viscosities and incoherent conductivity to obtain a dimensionless quantity (which we denote using a $*$ superscript:  e.g. $\sigma_0^*$). For the viscosities, we do so as follows: \begin{equation}
    \eta^* = \frac{\eta}{\gamma_{\mathrm{ee}} \langle v_{\mathrm{F}}^2\rangle \langle \mathsf{P}_x | \mathsf{P}_x \rangle},
\end{equation}
(bulk and rotational viscosity are defined using the same dividing factor), while for incoherent conductivity, \begin{equation}
    \sigma_0^* = \frac{\sigma_0}{\gamma_{\mathrm{ee}} \langle v_{\mathrm{F}}^2\rangle \langle \mathsf{n} | \mathsf{n} \rangle}.
\end{equation}
The normalized results are reported in Table \ref{tab:viscestimates}.

\begin{table}[]
    \centering
    \begin{tabular}{c|c|c|c}
           &  $E_F = 0.005$ eV & $E_F = 0.01$ eV & $E_F = 0.02$ eV\\
    \hline \hline
    $\eta^*$ &  0.3 & 0.4 & 0.25 \\ \hline
    $\tilde{\eta}^*$ & 0.3 & 0.3 & 0.1 \\ \hline
    $\zeta^*$ & 0.3 & 0.4 & 0.01 \\ \hline
    $\sigma_0^*$ & $0.6$ & $0.4$ & $0.1$
    \end{tabular}
    \caption{The dimensionless normalized viscosities and incoherent conductivity for the three Fermi surfaces.  The computation is a bit sensitive to the precise value of $T_*$ used, so we only report an estimate to one significant digit.  }
    \label{tab:viscestimates}
\end{table}

We find that the normalized incoherent conductivity is largest in Fermi surface \ref{fig:disconnectedFS}, consistent with the observation of diffusive transport in a regime which was hydrodynamic on other Fermi surfaces.   More interestingly, we see that $\sigma_0^*$, $\zeta^*$ and $\tilde\eta^*$ are all comparably large to $\eta^*\sim 0.3-0.4$ for Fermi surfaces \ref{fig:disconnectedFS} and \ref{fig:pocketFS}, while much smaller on the third Fermi surface (approximately a pair of triangles).  If the Fermi surface were a perfect circle, then we would have $\sigma_0^* = \zeta^* = \tilde\eta^* = 0$ in the low temperature limit \cite{Lucas_2018}:  the reason for this is that for an isotropic Fermi liquid, there is an approximate ``Galilean symmetry" as $T/T_{\mathrm{F}} \rightarrow 0$, whereby the difference between the current $|\mathsf{J}\rangle$ and momentum $|\mathsf{P}\rangle$ becomes subleading in $T/T_{\mathrm{F}}$, thereby making both $\zeta^*$ and $\sigma_0^*$ anomalously small in $T/T_{\mathrm{F}}$ \cite{dassarmalucas}.  Evidently, the polygonal Fermi surface is much closer to ``circular" than the locally disconnected Fermi surfaces.  Amusingly, this observation agrees with the prediction made in \cite{caleb} that polygonal Fermi surfaces would have very small $\tilde\eta^*$ and $\zeta^*$, albeit for a different reason than what was posited in that work (previously, it had been proposed that $\zeta^*$ is small due to the fluid arising out of a Fermi liquid, while $\tilde\eta$ would be small due to a hierarchy of scattering times).   

In particular, we find that the bulk viscosity $\zeta^*$ does \emph{not} need to be small in an anisotropic Fermi liquid.   From a formal (symmetry-based) perspective, this is assured because there will be infinitely many vectors $|\Phi\rangle$ in the kinetic formalism that transform in the trivial point group representation (thus possibly overlapping with the trace of the stress tensor $|\mathsf{T}_{kk}\rangle$), while only one of them (the density $|\mathsf{n}\rangle$) is a conserved mode which cannot relax.  An obvious example of such a mode in the two-triangle Fermi surface  \ref{fig:pocketFS} is to consider a fluctuation in which the inner Fermi surface shrinks while the outer one expands -- this can be achieved in such a way that there is no net change in number density.  

We can now use these results to better understand our earlier numerics.  In the previous section we found that the flow pattern for Fermi surface \ref{fig:disconnectedFS} with $\lee = 0.05w$, $\lei = 2w$ exhibited ohmic characteristics (two maxima in the current profile inside of the constriction), even though the length scales led to a more manifestly viscous regime in the other two Fermi surfaces.  And indeed, we see from Table \ref{tab:viscestimates} that $\sigma_0^*$ is somewhat larger for Fermi surface \ref{fig:disconnectedFS} than for the other two.  This gives a qualitative resolution to that observation, and demonstrates that careful determination of scattering lengths in imaging experiments will rely on a proper treatment of the microscopic Fermi surface geometry.

For completeness, we also remind the reader of the well-known fact that in an ohmic theory, the conductivity tensor must be isotropic in a theory with this point group.

\subsection{Experimental estimates}\label{sec:compare}
Let us now give some heuristic estimates of the relevant length scales in two-dimensional graphene-based electron liquids, mostly taken from experiments on monolayer and bilayer graphene.

One usually finds that at low temperature,  $\lei \sim 4 \; \mu \mathrm{m}$ is roughly temperature independent.  In contrast, up to logarithms in $T$ (which are not easily discerned experimentally anyway), we may estimate the scaling of $\lee$ by the formula
\begin{equation}
    \lee \sim \frac{\hbar v_{\mathrm{F}} E_{\mathrm{F}}}{T^2} \sim 1 \; \mu \mathrm{m} \times \left(\frac{30 \; \mathrm{K}}{T}\right)^2.
\end{equation} 
Thus, approximately one tunes from ballistic to viscous or ohmic regimes by changing $T$.  In the formula above, we have adjusted for the fact that the Fermi velocities in ABA graphene are on the order of $0.5 \times 10^6$ m/s, about half that of monolayer graphene; we have also assumed that $E_{\mathrm{F}}\approx 0.02$ eV, consistent with the polygonal Fermi surface \ref{fig:triangleFS}.  In fact, in ABA graphene, $T\sim 130$ K is comparable to the Fermi energy for Fermi surfaces \ref{fig:disconnectedFS} and \ref{fig:pocketFS}.  As a consequence, it may be more challenging to apply the Fermi liquid hydrodynamics described above for these two Fermi surfaces; one will also need to consider energy conservation, analogous to a Dirac fluid \cite{Crossno_2016,Gallagher2019}.  We anticipate the primary effect of this added conservation law is to enhance the ohmic signal relative to the viscous signal.

For the Fermi surface \ref{fig:triangleFS}, assuming a constriction of width $w=1\; \mu \mathrm{m}$ (which can easily be engineered in experiment), then the ballistic regime with $\lee = w$ will occur at around 30 K.  In contrast, an electron-electron scattering length of $\lee = 0.05w$ occurs at at $T \sim 130$ K.  These are easily achieved in experiments.   In any event, one can always cool down as low as possible in order to see a cleaner ballistic signal.

Recall that the $T_*$ we used in numerics was of order 3 K.  This means that the Fermi surface is much less sharply defined at the temperatures where we estimate viscous hydrodynamics will be easiest to observe.   This may lead to some quantitative changes in the predicted flow patterns; however, the numerical simulations needed to generate the plots are substantially more time-consuming.  It is straightforward but tedious to adjust our algorithm to more accurately include thermal effects, so we leave this generalization for another work.


\section{Comparison to model with cartoon Fermi surface}
\begin{figure}[t]
    \centering
    \begin{subfigure}[b]{0.45\textwidth}
        \includegraphics[width=\textwidth]{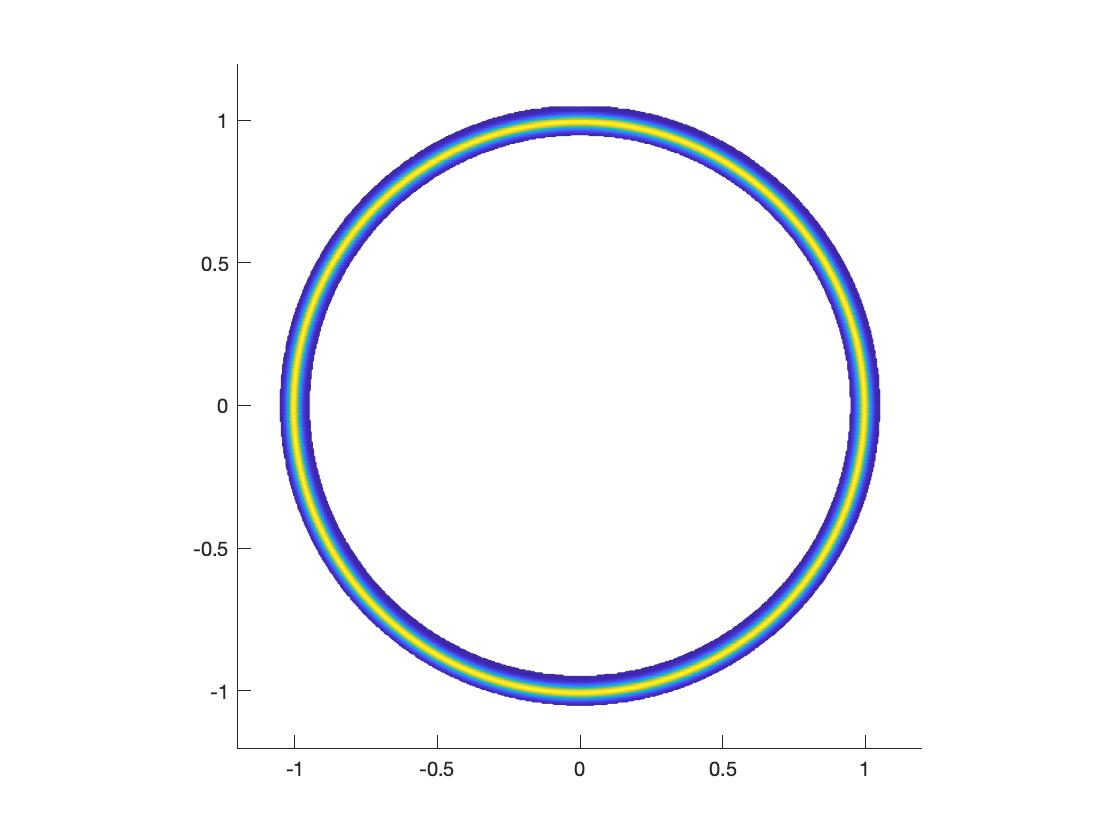}
        \label{fig:circleFS}
    \end{subfigure}
    \begin{subfigure}[b]{0.45\textwidth}
        \includegraphics[width=\textwidth]{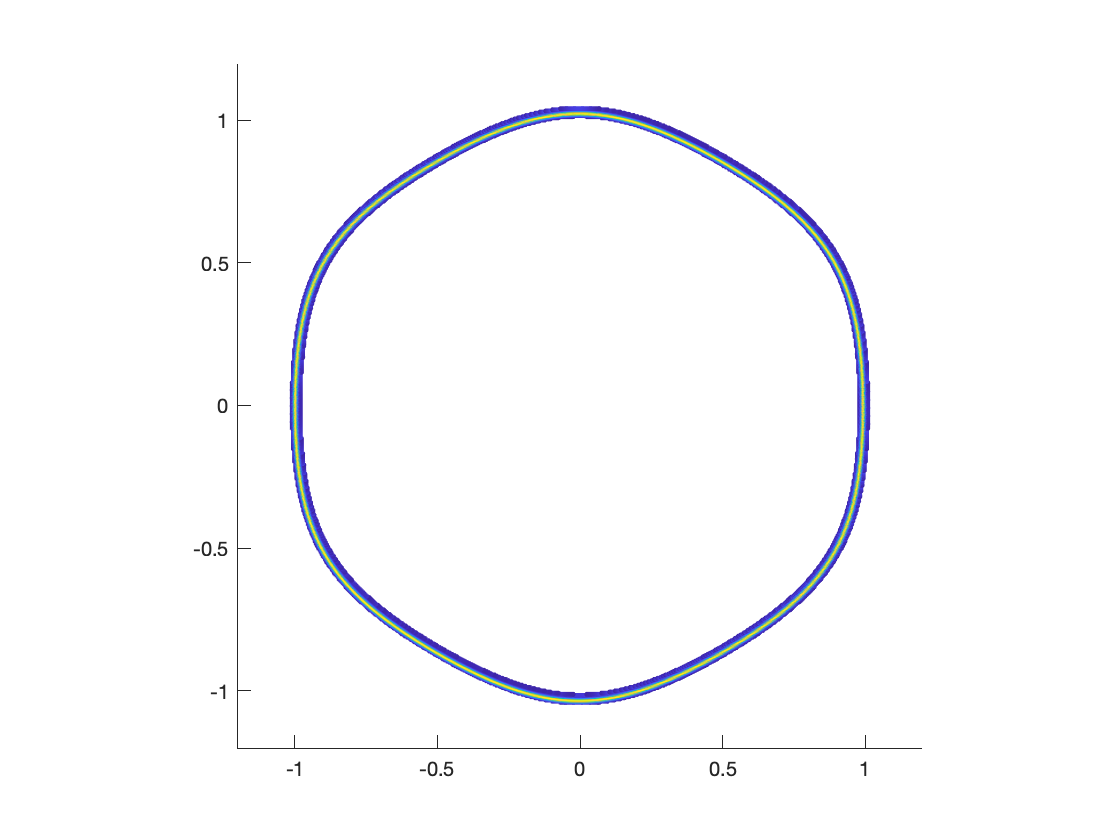}
        \label{fig:hexagonFS}
    \end{subfigure}
    \caption{Fermi surfaces for the band structures in Eq. \ref{eq:circ} (left) and Eq. \ref{eq:hex} (right).}
    \label{fig:cartoonfs}
\end{figure}

The qualitative features of the current profiles in ABA graphene can also be observed in simpler cartoon band structures. For comparison, we have performed an identical calculation to those described in the previous section, for two cartoon Fermi surfaces.  Firstly, as done in \cite{Guo_2017}, we calculate the current flow patterns for a circular Fermi surface (relevant for low density monolayer or bilayer graphene, or GaAs, e.g.):
\begin{equation}\label{eq:circ}
    \epsilon_{\Circle} = \alpha(p_x^2 + p_y^2).
\end{equation}
Secondly, we present a toy model for a hexagonal Fermi surface, found by constructing a polynomial which is explicitly invariant under the appropriate point group:
\begin{equation}\label{eq:hex}
    \epsilon_{\hexagon} = \beta\left[p_x^6 + \left(\frac{p_x}{2}-\frac{\sqrt{3}p_y}{2}\right)^6 +  \left(\frac{p_x}{2}+\frac{\sqrt{3}p_y}{2}\right)^6  \right].
\end{equation}
The Fermi surfaces for these two dispersion relations are shown in Fig. \ref{fig:cartoonfs}. The computation of $\Sigma(\mathbf{k})$ for these Fermi surfaces again follows from the procedure outlined in Section \ref{sec:kinetictheory}. 

\begin{figure}[t]
    \centering
    \begin{subfigure}{0.32\textwidth}
        \caption{$\ell_{ee} = 0.05w$, $\ell_{ei} = 0.2w$}
        \vspace{-7pt}
        \includegraphics[width=\textwidth]{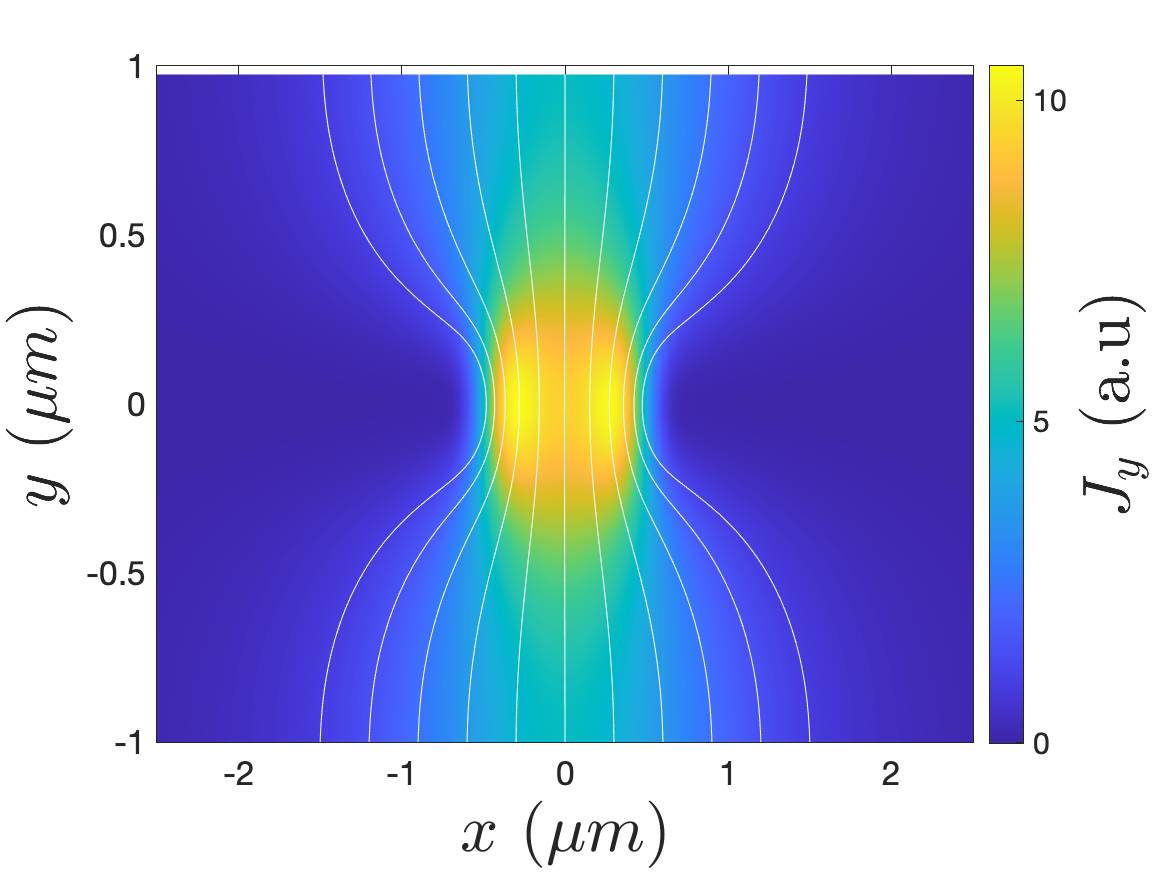}    
        \label{fig:circle_ohmic}
    \end{subfigure}
    \begin{subfigure}{0.32\textwidth}
        \caption{$\ell_{ee} = 0.05w$, $\ell_{ei} = 2w$}
        \vspace{-7pt}
        \includegraphics[width=\textwidth]{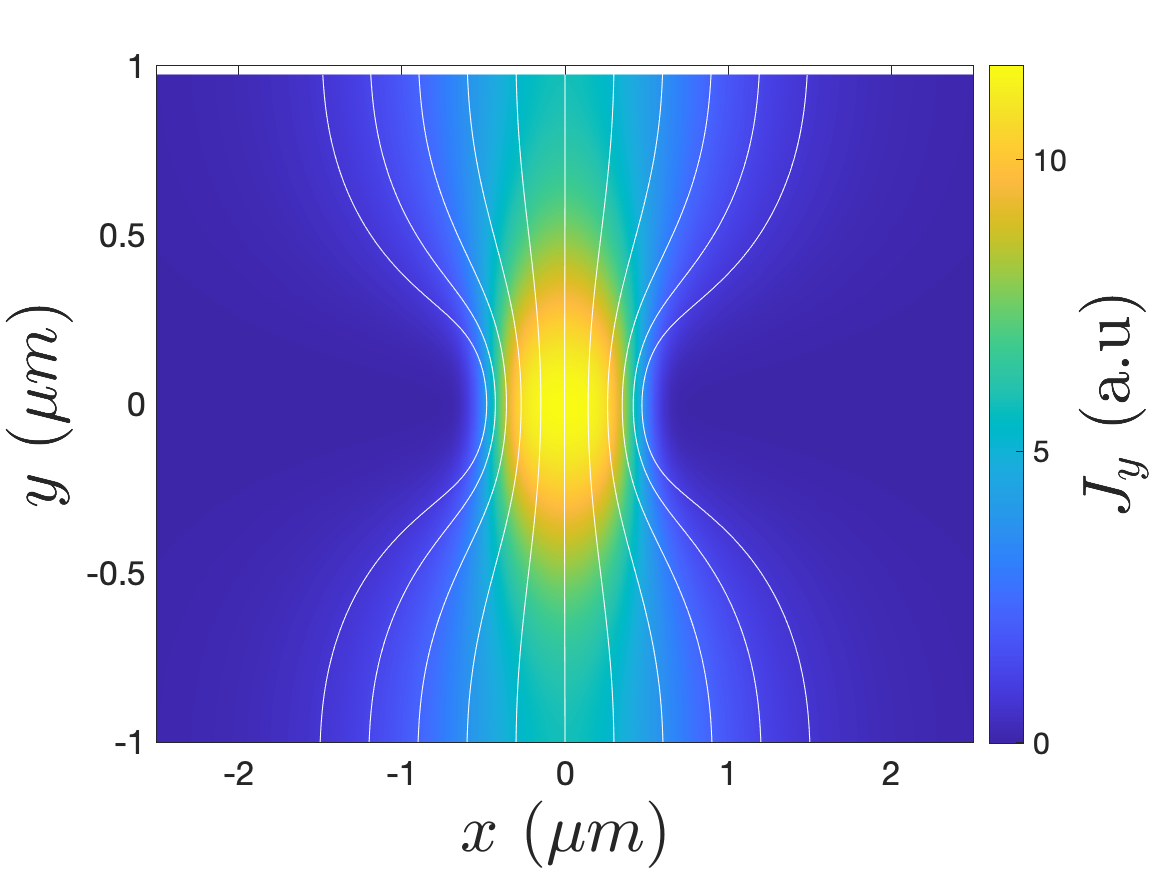}
        \label{fig:circle_hydro}
    \end{subfigure}
    \begin{subfigure}{0.32\textwidth}
        \caption{$\ell_{ee} = w$, $\ell_{ei} = 2w$}
        \vspace{-7pt}
        \includegraphics[width=\textwidth]{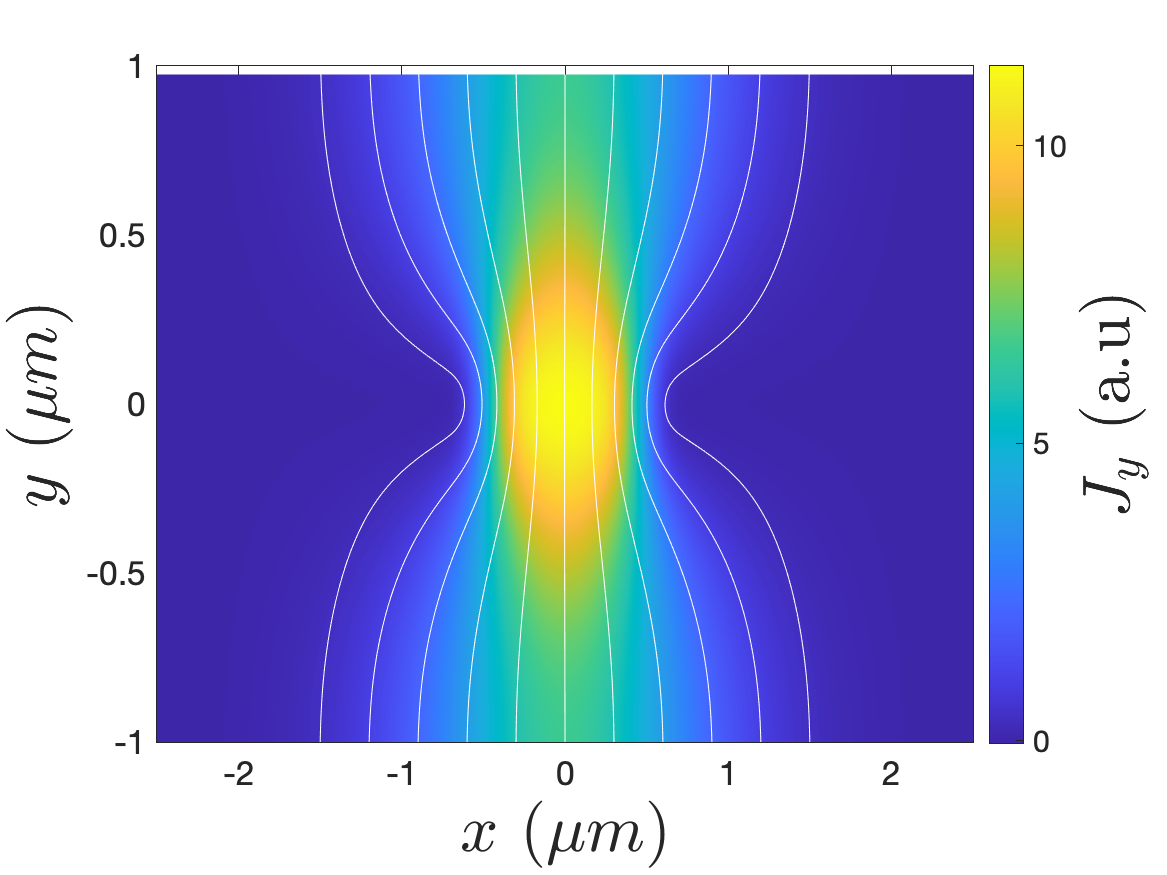}
        \label{fig:circle_ballistic}
    \end{subfigure}
    \caption{Flow patterns for the $\lee = 0.05w$, $\lei = 0.2w$ (left), $\lee = 0.05w$, $\lei = 2w$ (middle), and $\lee = w$, $\lei = 2w$ (right) for the circular Fermi surface. }
    \label{fig:flow_circle}
\end{figure}

\begin{figure}[t]
    \centering
    \begin{subfigure}{0.32\textwidth}
        \caption{$\ell_{ee} = 0.05w$, $\ell_{ei} = 0.2w$}
        \vspace{-7pt}
        \includegraphics[width=\textwidth]{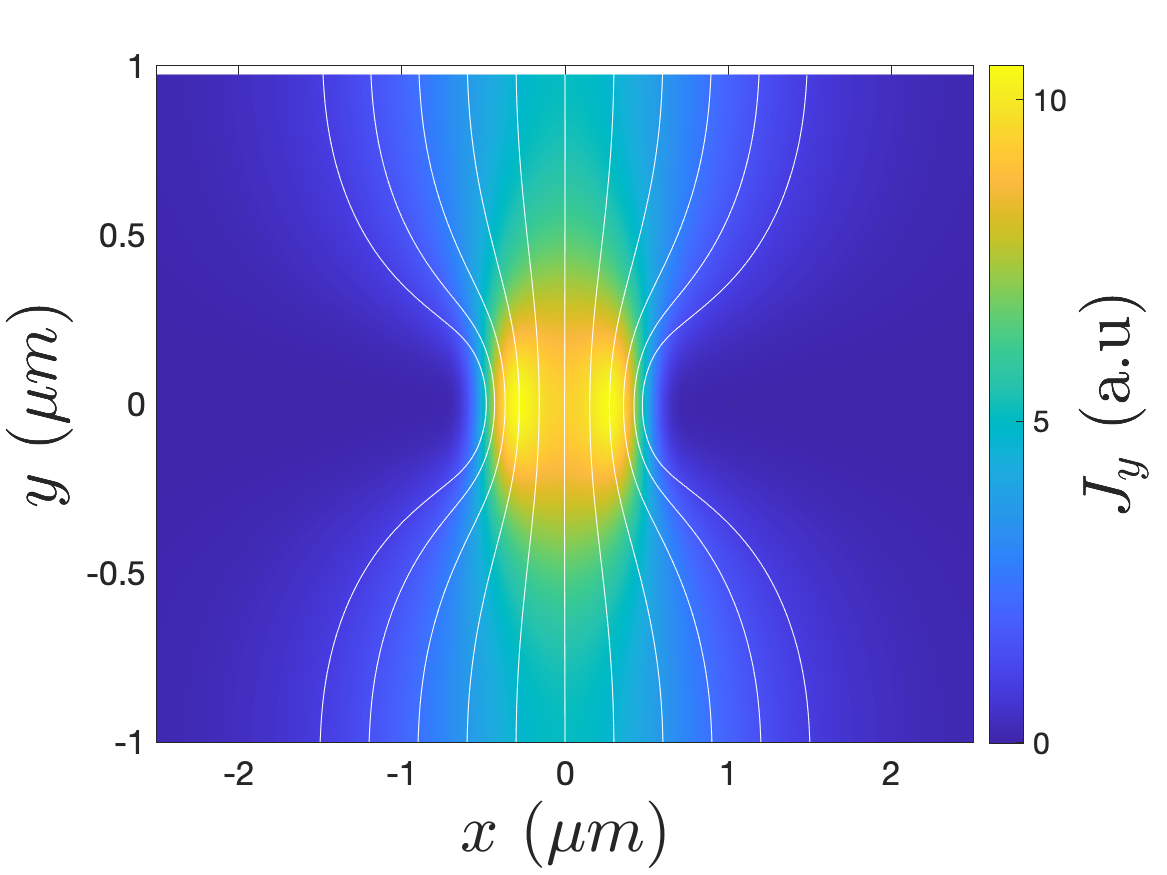}   
        \label{fig:hexagon_ohmic}
    \end{subfigure}
    \begin{subfigure}{0.32\textwidth}
        \caption{$\ell_{ee} = 0.05w$, $\ell_{ei} = 2w$}
        \vspace{-7pt}
        \includegraphics[width=\textwidth]{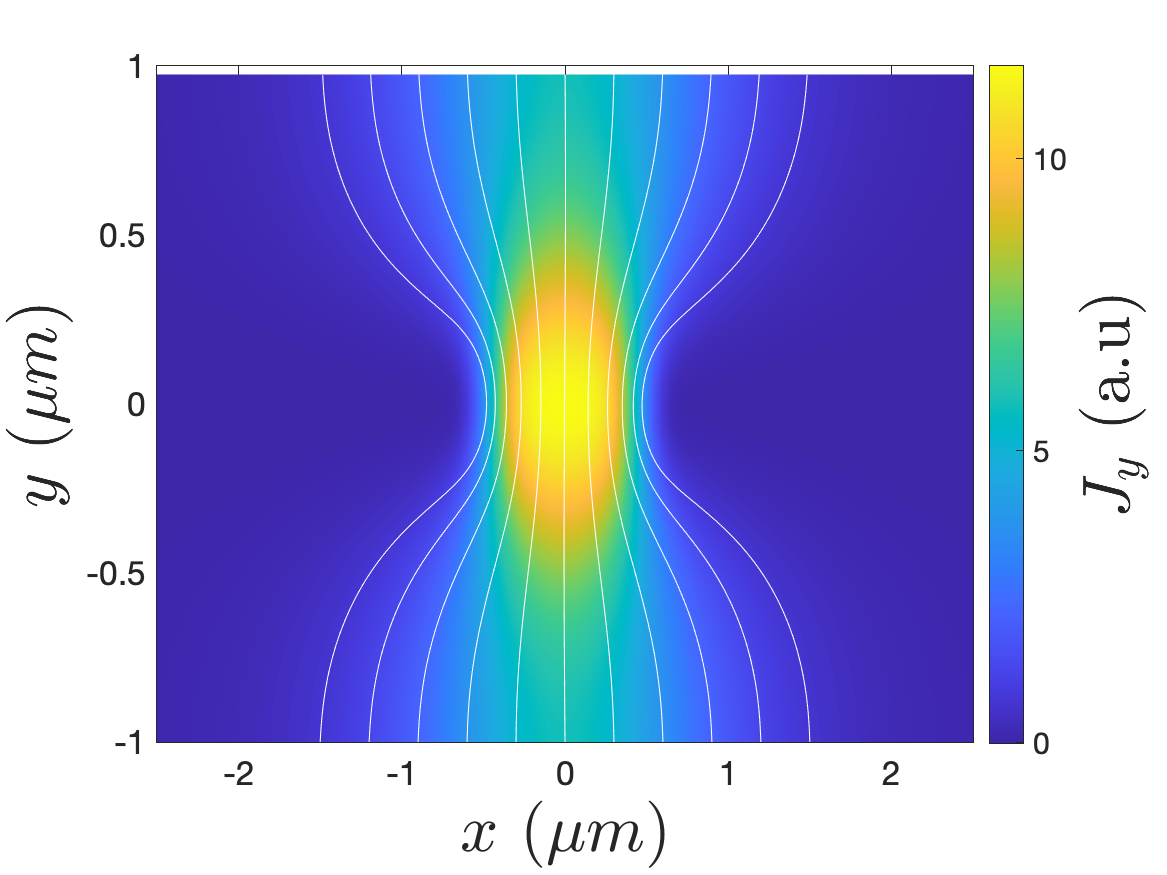}
        \label{fig:hexagon_hydro}
    \end{subfigure}
    \begin{subfigure}{0.32\textwidth}
        \caption{$\ell_{ee} = w$, $\ell_{ei} = 2w$}
        \vspace{-7pt}
        \includegraphics[width=\textwidth]{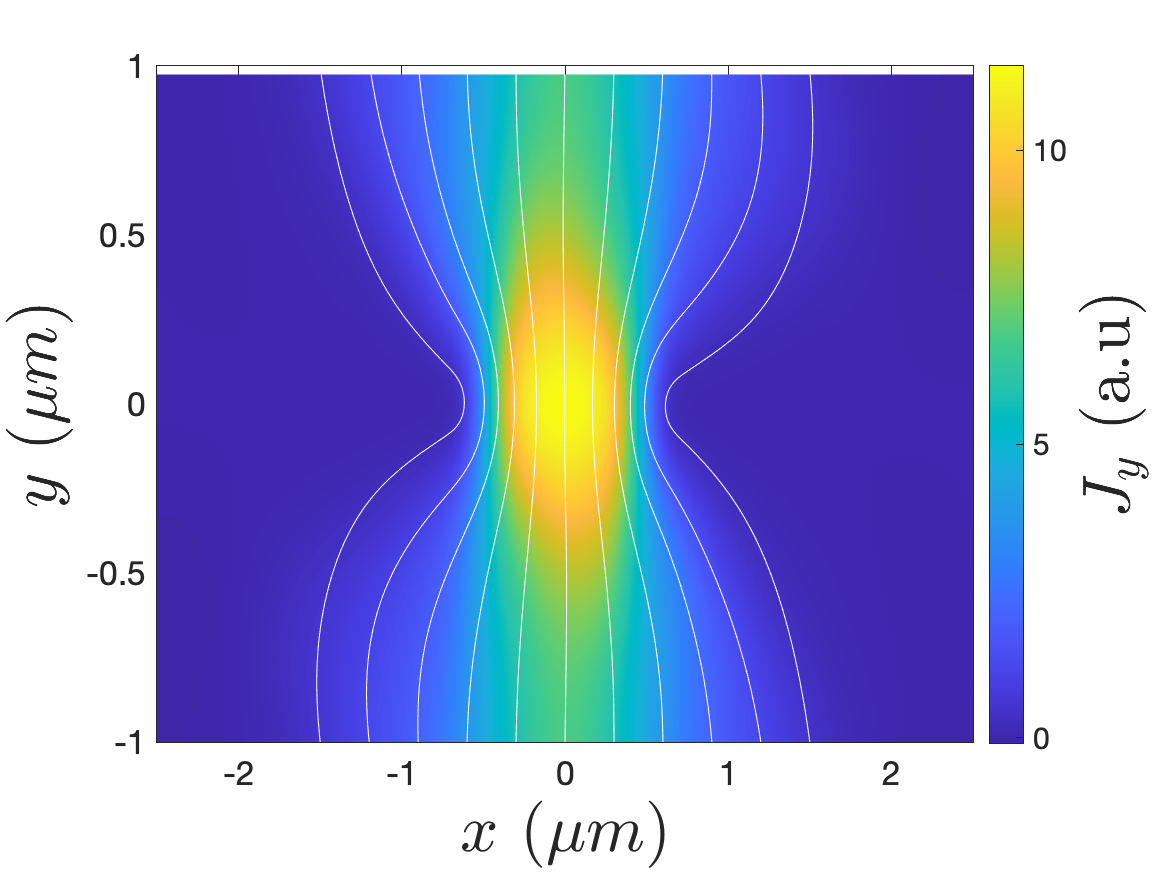}
        \label{fig:hexagon_ballistic}
    \end{subfigure}
    \caption{Flow patterns for $\lee = 0.05w$, $\lei = 0.2w$ (left), $\lee = 0.05w$, $\lei = 2w$ (middle), and $\lee = w$, $\lei = 2w$ (right) for the hexagonal Fermi surface. In each plot, the Fermi surface is rotated $0.35$ radians relative to the constriction.  }
    \label{fig:flow_hex}
\end{figure}

The resulting flow patterns are shown in Fig. \ref{fig:flow_circle} for the circular Fermi surface and in Fig. \ref{fig:flow_hex} for the hexagonal Fermi surface. As before, the length scales are chosen to obtain flow patterns in the ohmic, viscous, and ballistic regimes. We use the same length scales as in Section \ref{sec:abaflow}: $\lee = 0.05w$, $\lei = 0.2w$ for the ohmic regime, $\lee = 0.05w$, $\lei = 2w$ for the hydrodynamic regime, and $\lee = w$, $\lei = 2w$ for the ballistic regime. These correspond to the left, middle, and right plots of Figs. \ref{fig:flow_circle} and \ref{fig:flow_hex}. Cross sections at $y=0$ and $y=-0.5$ $\mu$m are shown in the left and middle plots of Fig. \ref{fig:currentxc_polygon}.

The flow patterns in the constriction geometry for the circular Fermi surface were previously discussed in \cite{jenkins2020imaging} in the context of monolayer graphene. Because the Fermi surface is isotropic, the asymmetric flow pattern and tilted streamlines which distinguish ballistic from hydrodynamic transport in ABA graphene are absent here. While distinguishing ballistic from hydrodynamic transport is difficult on the basis of the flow patterns in Fig. \ref{fig:flow_circle} alone, it was noted in \cite{jenkins2020imaging} that the shape of the graph of $J_y$ vs. $x$ at $y=0$ can be used to differentiate between them. 

The flow patterns for the hexagonal Fermi surface resemble those of ABA graphene. In contrast to the circular Fermi surface, there is a clear qualitative signature of ballistic transport, namely the asymmetry of the flow pattern. We note, however, that the signature of ballistic transport for the hexagonal Fermi surface is not as prominent as it is in ABA graphene despite the electron-electron and electron-impurity length scales being the same. This appears to be due to an increase in curvature near the corners of the hexagonal Fermi surface, which causes the Fermi velocities on an edge to deviate slightly from being parallel. This causes the streamlines to be less straight than in the case of ABA graphene. Nevertheless, the off-center peak in the $y=-0.5$ $\mu$m current cross section, shown in the middle panel of Fig. \ref{fig:currentxc_polygon}, provides a complementary unambiguous indicator of ballistic transport.

\begin{figure}[t]
    \centering
    \begin{subfigure}{0.32\textwidth}
        \caption{$\theta = 0$}
        \vspace{-7pt}
        \includegraphics[width=\textwidth]{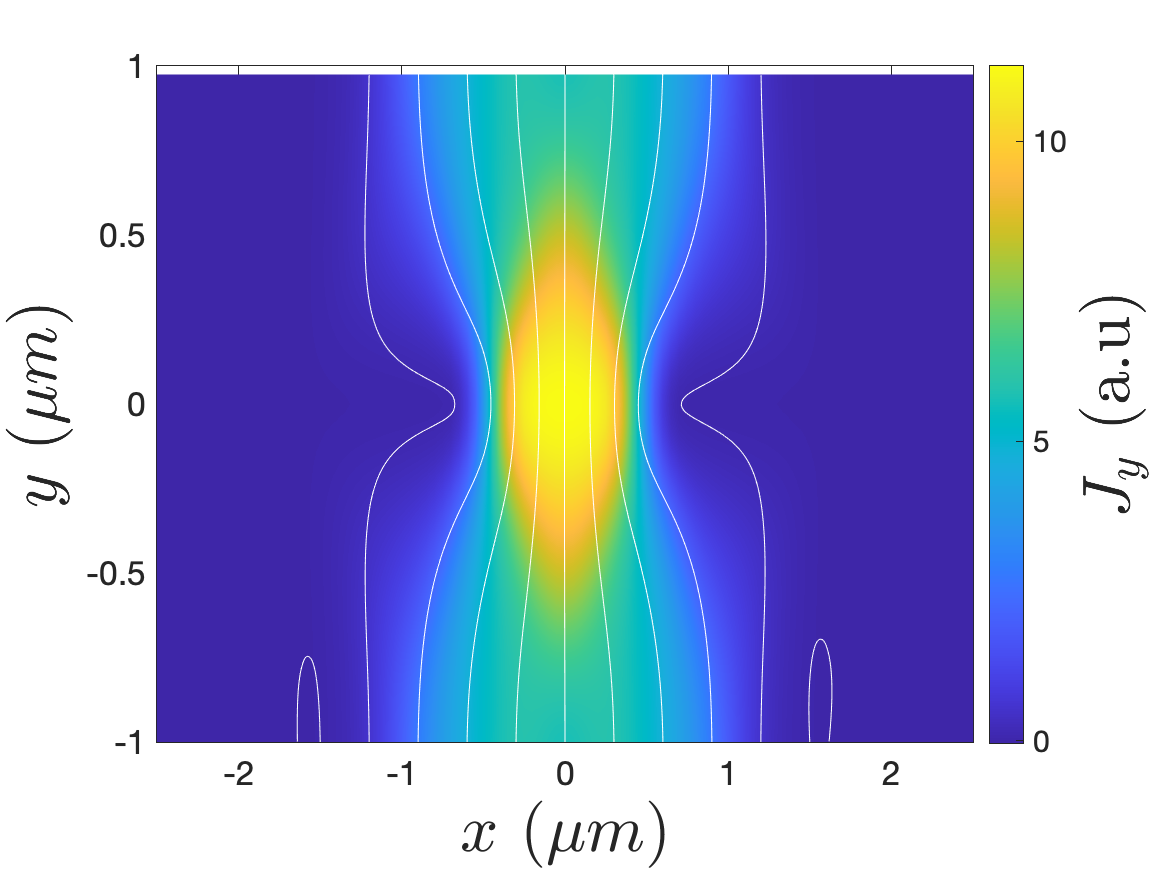}
        \label{fig:hexagon_notilt}
    \end{subfigure}
    \begin{subfigure}{0.32\textwidth}
        \caption{$\theta = \pi/12$}
        \vspace{-7pt}
        \includegraphics[width=\textwidth]{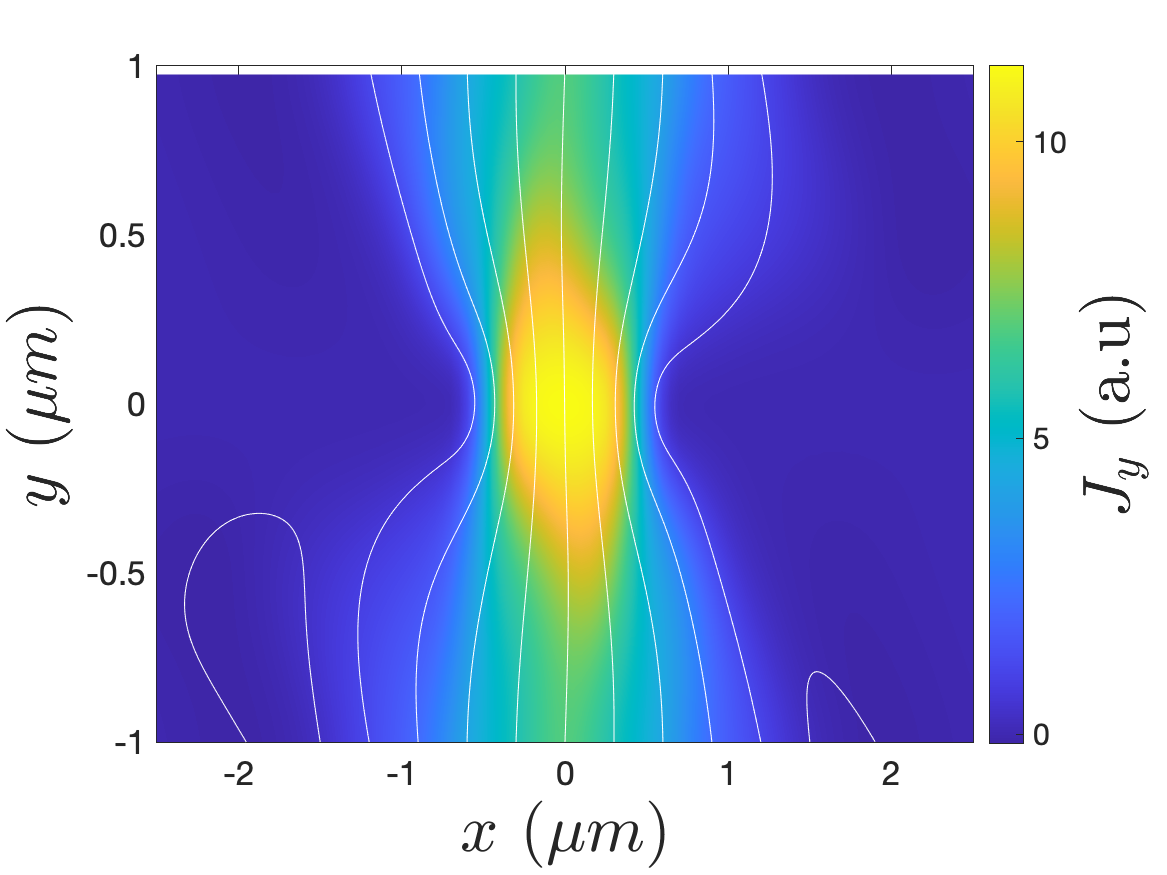}
        \label{fig:hexagon_tilt}
    \end{subfigure}
    \begin{subfigure}{0.32\textwidth}
        \caption{$\theta = \pi/6$}
        \vspace{-7pt}
        \includegraphics[width=\textwidth]{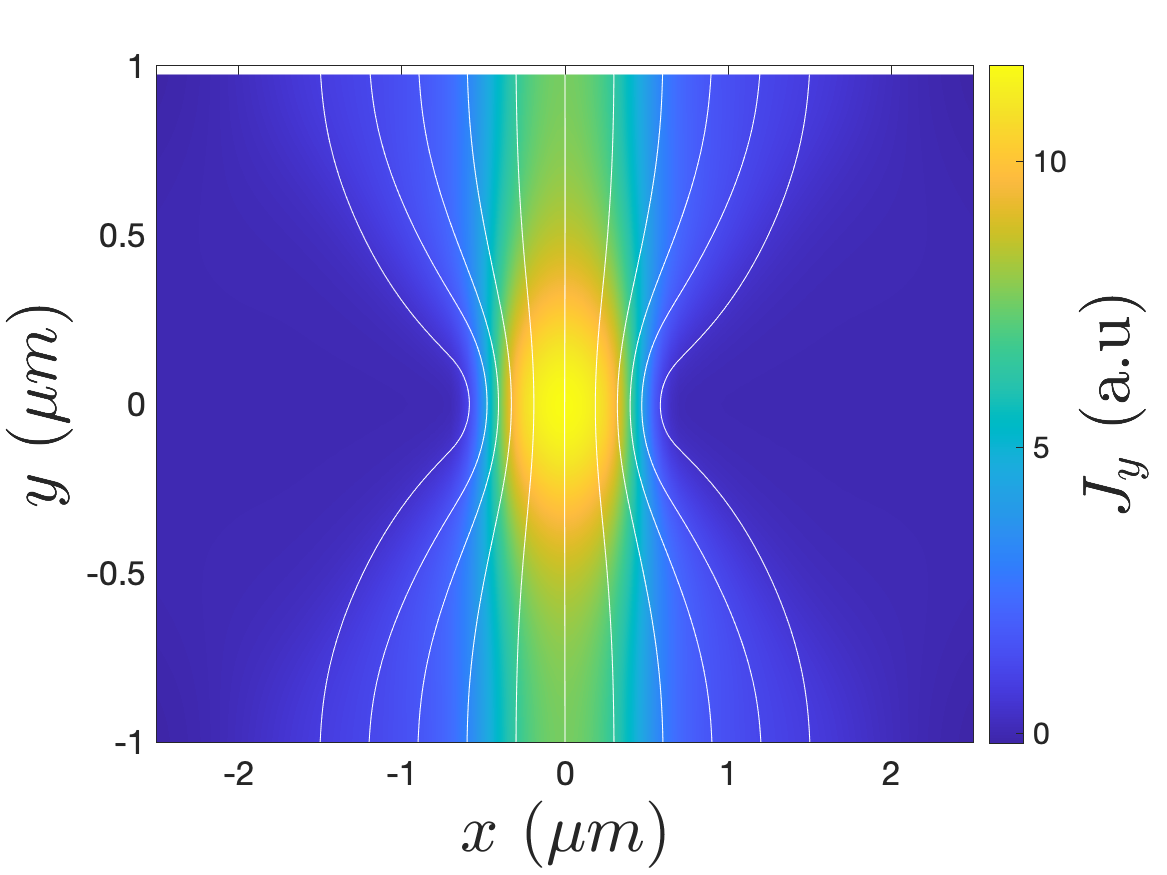}
        \label{fig:hexagon_symtilt}
    \end{subfigure}
    \caption{Ballistic flow patterns ($\lee = 2w$, $\lei = 4w$) for the hexagonal Fermi surface rotated zero (left), $\pi/12$ (middle), and $\pi/6$ radians relative to the constriction.}
    \label{fig:ballistic_angles}
\end{figure}

\begin{center}
    \begin{figure}
        \centering
        \begin{tabular}{c c c}
        \includegraphics[width=0.33\textwidth]{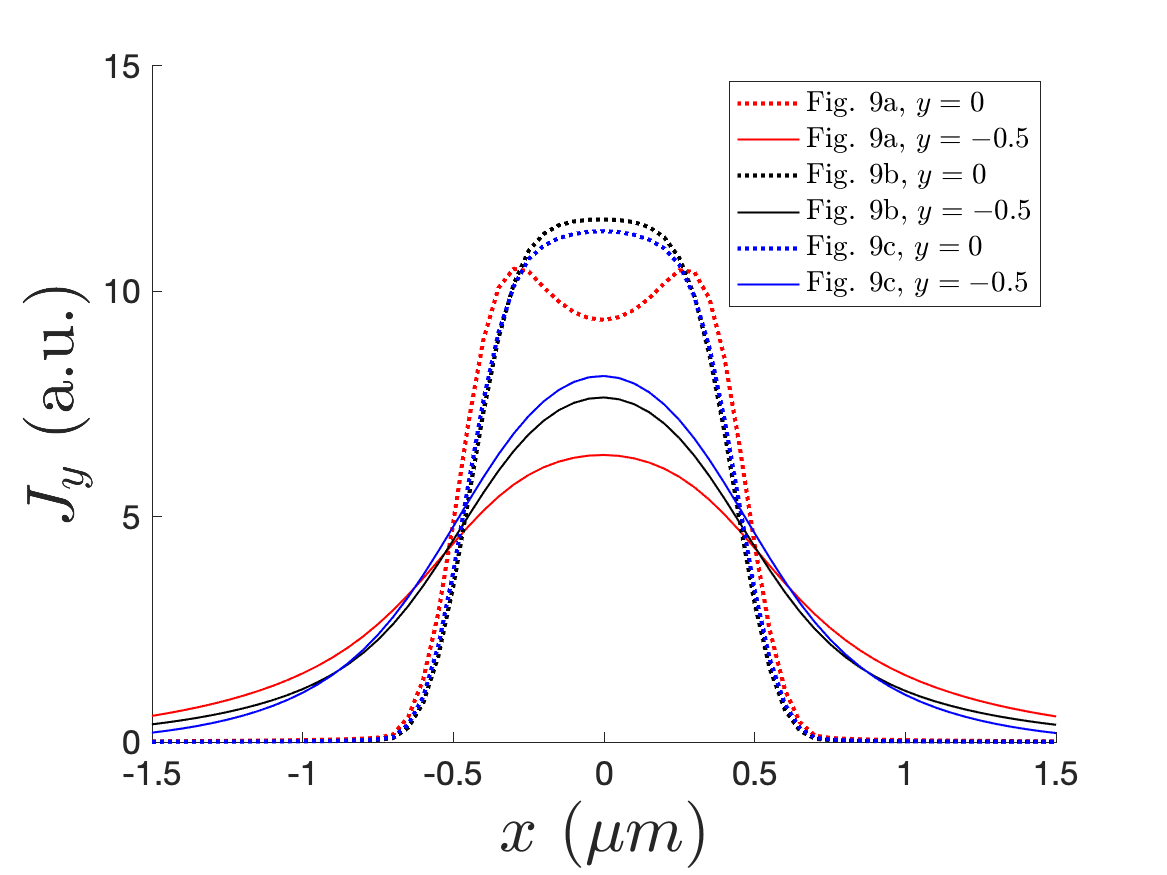} &
        \includegraphics[width=0.33\textwidth]{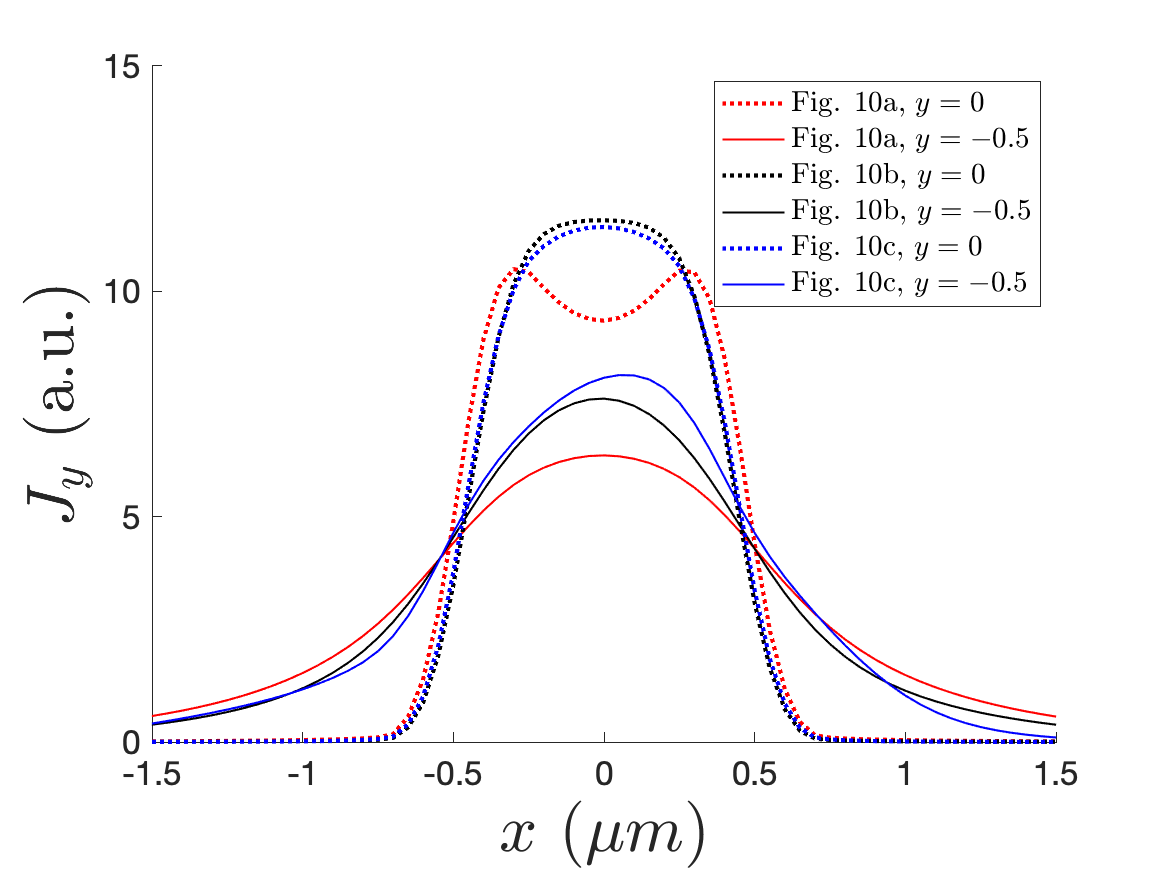} &
        \includegraphics[width=0.33\textwidth]{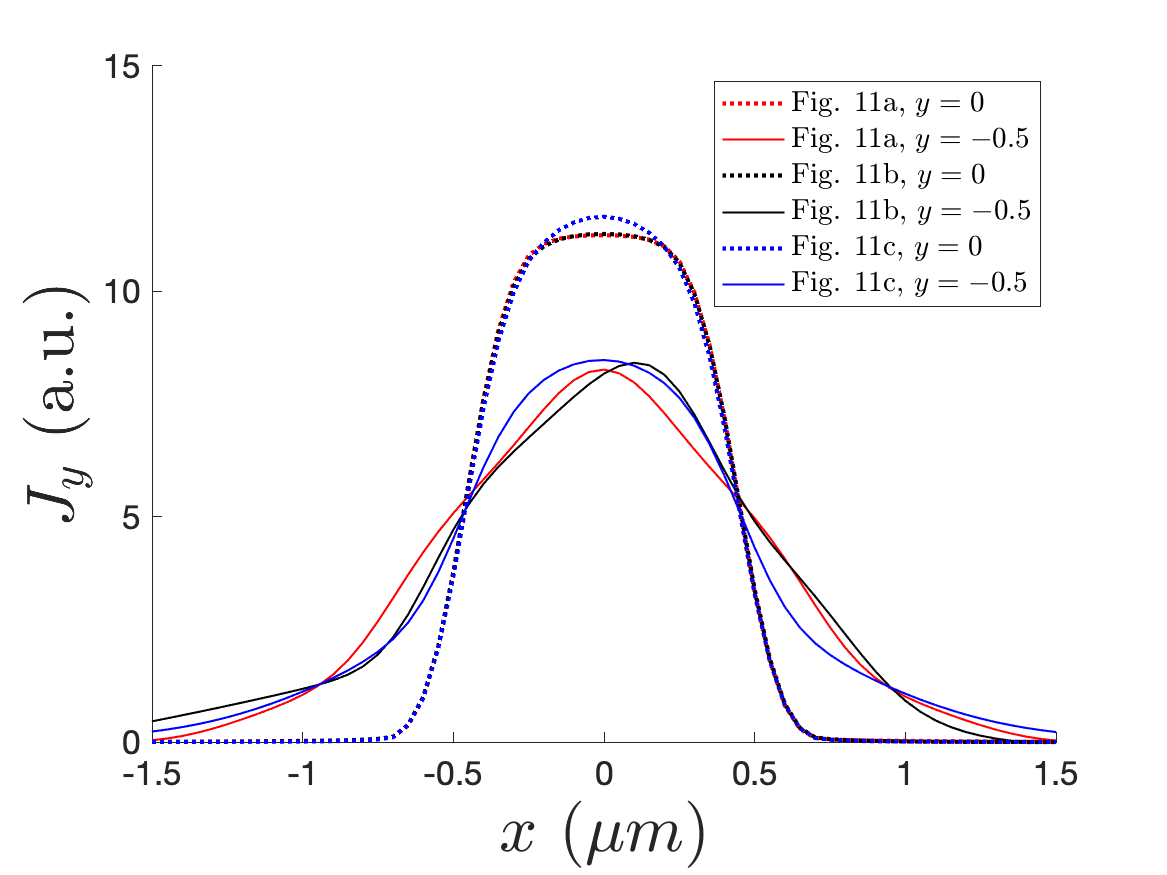}
        \end{tabular}
        \caption{Flow profiles $J_y$ vs. $x$ through the center of the slit ($y=0$) and through $y=-0.5 \; \mu\mathrm{m}$ for the flow patterns in Fig. \ref{fig:flow_circle} (left), \ref{fig:flow_hex} (middle), and \ref{fig:ballistic_angles} (right).}
        \label{fig:currentxc_polygon}
    \end{figure}
\end{center}

Finally, we comment on the importance of the relative orientation between the Fermi surface and the constriction. The relative orientation has no effect on the flow pattern within the ohmic or hydrodynamic regimes. In the ballistic regime, however, relative orientation is crucial to seeing the asymmetry of the current and streamlines. The effect is illustrated in Figure \ref{fig:ballistic_angles}. For $\theta = 0$ and $\theta = \pi/6$, the Fermi surface is reflection symmetric across the $y$-axis, and the resulting flow pattern is as well. For $\theta = \pi/12$, the asymmetry is manifest and ballistic transport can be easily distinguished. These features can also be seen in the right plot of Fig. \ref{fig:currentxc_polygon}, where the current cross section through $y=-0.5$ $\mu$m has an off-center peak only when $\theta = \pi/12$, and a symmetric peak when $\theta=0$ and $\theta = \pi/6$.

\section{Conclusion}
We have demonstrated that local imaging methods such as nitrogen vacancy center magnetometry can, in theory, distinguish qualitatively between ohmic, viscous and ballistic flow patterns using only a single snapshot of current flow. To summarize, the key distinguishing features of each regime are as follows: ohmic flows are symmetric, with two peaks near the constriction boundaries; viscous flows are symmetric with a single central peak; ballistic flows are asymmetric with visible current streaks, provided the Fermi surface is tilted and sufficiently anisotropic.  Although we believe ABA trilayer graphene is a particularly appealing material platform in which to look for these phenomena, our general framework is certainly more broadly applicable.  The framework we have outlined, which treats the microscopic band structure carefully while using heuristic models for both kinetic boundary conditions and precise collision integrals, may be a useful compromise for near-term comparison between theory and experiment.

In principle, the methods of Section \ref{sec:method} allow one to study non-quasiparticle systems \cite{huang2021fingerprints} using the same methods as we will use below to study Fermi liquids with well defined quasiparticles.  The key limitation of this approach is that we have used ``mystery" boundary conditions on the microscopic theory, by assuming the simple form of the induced electric field.  It has been noted \cite{Pershoguba2020} that even in complicated devices, the boundary conditions could play a qualitatively important role in determining flow patterns.  This is an important limitation of our approach that we hope to return to in a future work.  

\section*{Acknowledgements}

We are especially grateful to Andrea Young for alerting us to the literature on ABA graphene. MQ was supported by the National Defense Science and Engineering Graduate Fellowship (NDSEG) program. AL was supported by the Gordon and Betty Moore Foundation's EPiQS Initiative via Grant GBMF10279, and by a Research Fellowship from the Alfred P. Sloan Foundation under Grant FG-2020-13795. 

\begin{appendix}



\end{appendix}

\bibliography{thebib}

\end{document}